\documentclass[useAMS,usenatbib]{mn2e}

\usepackage[dvips]{graphicx}
\usepackage{lscape}
\usepackage{deluxetable}
\usepackage{amsmath,amsfonts,amssymb}
\usepackage[dvips]{color}
\usepackage[bottom]{footmisc}

\def\deg {^{\circ} }
\def\sqdeg {\,deg$^2$}
\def\arcsec {\,arcsec}

\title[ATLAS Third Data Release]{ATLAS -- I. Third Release of 1.4~GHz Mosaics and Component Catalogues}

\author[T.~M.~O.~Franzen et al.]{T.~M.~O.~Franzen$^{1,2}$ \thanks{Email: thomas.franzen@curtin.edu.au},
J.~K.~Banfield$^{3,11}$, C.~A.~Hales$^{4,5}$, A.~Hopkins$^{6}$, R.~P.~Norris$^{2}$,  
\newauthor N.~Seymour$^{1,2}$, K.~E.~Chow$^{2}$, A.~Herzog$^{7,8,2}$, M.~T.~Huynh$^{9}$, E.~Lenc$^{10,11}$, M.~Y.~Mao$^{4}$,
\newauthor E.~Middelberg$^{7}$\\
$^{1}$International Centre for Radio Astronomy Research (ICRAR), Curtin University, Perth, Australia \\
$^{2}$CSIRO Australia Telescope National Facility, PO Box 76, Epping, NSW, 1710, Australia\\
$^{3}$Research School of Astronomy and Astrophysics, Australian National University, Weston Creek, ACT 2611, Australia\\
$^{4}$National Radio Astronomy Observatory, P.O. Box 0, Socorro, NM 87801, USA\\
$^{5}$Jansky Fellow of the National Radio Astronomy Observatory\\
$^{6}$Australian Astronomical Observatory, PO Box 915, North Ryde, NSW 1670, Australia\\
$^{7}$Astronomisches Institut, Ruhr-Universit{\"a}t Bochum, 44801 Bochum, Germany\\
$^{8}$Macquarie University Research Centre for Astronomy, Astrophysics, and Astrophotonics, North Ryde, NSW 2109, Australia\\
$^{9}$International Centre for Radio Astronomy Research, University of Western Australia, Crawley, WA 6009, Australia\\
$^{10}$Sydney Institute for Astronomy, School of Physics, University of Sydney, NSW 2006, Australia\\
$^{11}$ARC Centre of Excellence for All-Sky Astrophysics (CAASTRO)}

\begin{document}

\date{Accepted year month day. Received year month day; in original form year month day}

\pagerange{\pageref{firstpage}--\pageref{lastpage}} \pubyear{2013}

\maketitle

\label{firstpage}

\begin{abstract}
We present the third data release from the Australia Telescope Large Area Survey (ATLAS).  These data combine the observations at $1.4\,$GHz before and after upgrades to the Australia Telescope Compact Array reaching a sensitivity of $14\,\mu$Jy beam$^{-1}$ in 3.6\,\sqdeg\ over the {\it Chandra} Deep Field South (CDFS) and of $17\,\mu$Jy beam$^{-1}$ in 2.7\,\sqdeg\ over the European Large Area {\it ISO} Survey South 1 (ELAIS-S1). We used a variety of array configurations to maximise the $uv$ coverage resulting in a resolution of 16 by 7~arcsec in CDFS and of 12 by 8~arcsec in ELAIS-S1. After correcting for peak bias and bandwidth smearing, we find a total of 3034 radio source components above $5\sigma$ in CDFS, of which 514 (17 per cent) are considered to be extended. The number of components detected above $5\sigma$ in ELAIS-S1 is 2084, of which 392 (19 per cent) are classified as extended. The catalogues include reliable spectral indices ($\Delta \alpha < 0.2$) between 1.40 and 1.71~GHz for $\sim 350$ of the brightest components.
\end{abstract}


\section{Introduction}\label{Introduction}

Large multiwavelength surveys are indispensable for understanding galaxy formation and evolution. Radio wavelengths are valuable in providing an obscuration-independent tracer of star formation and active galactic nucleus (AGN) activity. However, most radio surveys so far have either covered only small areas, thus suffering from sample and cosmic variance, and missing intrinsically rare objects, or cover wide areas but are relatively insensitive, and therefore miss the most active epochs of galaxy formation. In particular, most wide radio surveys have not had sufficient sensitivity to detect normal star formation activity in any but the most nearby galaxies, limiting their ability to contribute to our understanding of the cosmic evolution of galaxies. Another important requisite to maximising the astrophysical value of a survey at any wavelength is to maximise the overlap with other wavelengths.

Here we present the third data release of the Australia Telescope Large Area Survey (ATLAS), which covers $\sim 6$\,\sqdeg\ to an rms depth of $\sim 15\,\mu$Jy beam$^{-1}$. The  ATLAS survey area consists of two regions centred on the {\it Chandra} Deep Field South \citep[CDFS;][]{Giacconi2001} and the European Large Area {\it ISO} Survey - South 1 \citep[ELAIS-S1;][]{Oliver2000}. These two fields were carefully chosen to coincide with areas imaged by the Spitzer Wide-area Infrared Extragalactic Survey \citep[SWIRE;][]{Lonsdale2003} program, so that infrared and optical data are available for most of the radio objects. They also encompass the well-studied Great Observatories Origins Deep Survey (GOODS) field in CDFS \citep{Giavalisco2004}.

These two areas have since been the target of many other deep multi-wavelength surveys, such as the 4~Ms X-ray survey from \textit{Chandra} \citep{xue2011}, the Spitzer Extragalactic Representative Volume Survey \citep[SERVS;][]{Mauduit2012}, the VISTA Deep Extragalactic Observations \citep[VIDEO;][]{Jarvis2013} survey and the Herschel Multi-tiered Extragalactic Survey \citep[HERMES;][]{Oliver2012} as well as spectroscopic and photometric redshift surveys, including the PRIsm MUlti-object Survey \citep[PRIMUS;][]{Coil2011,Cool2013}, the FourStar Galaxy Evolution Survey \citep[zFOURGE;][]{Spitler2012}, the ATLAS spectroscopy program
 \citep{Mao2012} and the new OzDES spectroscopy program (Lidman et al., in preparation), making them some of the best-studied fields in the sky. 

The first ATLAS data release \citep[DR1:][]{Norris2006,Middelberg2008} surveyed these two fields in CDFS and ELAIS-S1 to a typical rms of 36 and $29\,\mu$Jy beam$^{-1}$ respectively. The second data release \citep[DR2:][]{Hales2014a,Hales2014b} surveyed them to a typical rms of $30\,\mu$Jy beam$^{-1}$, addressing a number of sources of systematic error in DR1 and for the first time presenting ATLAS polarisation results. Here in DR3 we report a further improvement to respective rms sensitivities of 14 and $17\,\mu$Jy beam$^{-1}$.

Minimising the rms noise is critical, because deep radio surveys such as ATLAS probe flux densities approaching the point where star-forming galaxies start to dominate the radio sky. Surveys of radio sources with flux densities greater than 1~mJy are typically dominated by AGNs, but sources at lower flux densities are increasingly driven by star formation activity \citep[e.g.][]{Seymour2008, Smolcic2008, Bonzini2013}. Since star-forming galaxies also dominate non-radio surveys, the fraction of optical/IR galaxies detected at radio wavelengths rises sharply with decreasing flux density, yielding an extinction-free measure of star formation rate.

The key science goals of ATLAS are:
\begin{itemize}
\item To determine the relative contribution of starbursts and AGN to the overall energy density of the universe, and the relationship between AGN and star-forming activity. Particularly interesting are composite galaxies in which a radio AGN lies buried within a host galaxy whose optical/infrared spectrum or SED appears to be that of a star-forming galaxy. 
\item To test whether the far-infrared-radio correlation changes with redshift or with other galaxy properties. Once calibrated, this correlation will be an effective method to measure the star formation history of the Universe.
\item To trace the radio luminosity function to a high redshift (z $\sim$ 1) for moderate-power sources, and measure the differential 20~cm source count to a flux density limit of $\sim 100$~$\mu\mathrm{Jy}$ to a high precision.
\item To explore regions of parameter space, corresponding to a large area of sky surveyed at high sensitivity at radio, mid-infrared, and far-infrared wavelengths, to discover rare but important objects, such as short-lived phases in galaxy evolution.
\end{itemize}

A supplementary but important goal is to act as a pathfinder for 
the Evolutionary Map of the Universe \citep[EMU;][]{Norris2011} survey,  which will use the new Australian SKA Pathfinder \cite[ASKAP;][]{Johnston2007, Johnston2008, Deboer2009} telescope to make a deep ($10\,\mu$Jy beam$^{-1}$ rms) radio continuum survey of the entire Southern Sky, extending as far North as $+30\deg$. EMU will cover roughly the same fraction (75\%) of the sky as the benchmark NVSS survey \citep{Condon1998}, but will be 45 times more sensitive, and will have an angular resolution (10~arcsec) 4.5 times higher. EMU is expected to generate a catalogue of about 70 million galaxies, compared to the $\sim$ 2.5 million sources currently known at all radio frequencies. Since EMU will have a similar resolution to ATLAS, we are using ATLAS to test many of the technical and scientific processes for EMU. In addition, we will use optical spectroscopy of ATLAS galaxies to train the photometric redshift algorithms for EMU.

Throughout this paper we define a radio `component' as a discrete region of radio emission identified in the source extraction process. We define a radio `source' as one or more radio components that appear to be physically connected to one host galaxy. Thus, we count a classical triple radio-loud source as being a radio source consisting of three radio components, but count a pair of interacting starburst galaxies as being two sources, each with one radio component (providing of course that the angular separation between the starburst galaxies is large enough for them to be resolved).

This paper (Paper I) is primarily concerned with describing the survey and presenting the component catalogue. Paper II (Banfield et al. 2015, in preparation) will extract the component counts and explore the distribution of spectral indices, and Paper III (Norris et al. 2015, in preparation)  will group the components into sources with optical/infrared identifications. Subsequent papers in the series will address the individual science goals of ATLAS.

This paper is organised as follows. Section~\ref{Observations, Calibration, and Imaging} describes the observations, calibration and imaging, and Section~\ref{Correction of Final Mosaics} details the corrections made to the final image prior to component  extraction. Section~\ref{Source Component Catalog} describes the component extraction process and presents the ATLAS DR3 component catalogue.  
We summarise our results in Section~\ref{Summary}.

Throughout this paper we assume a Hubble constant of $71\,{\rm km}\,{\rm s}^{-1}{\rm Mpc}^{-1}$, and matter and cosmological constant density parameters of $\Omega_{\rm M}=0.27$ and $\Omega_{\rm \Lambda}=0.73$, and assume the convention for spectral index, $\alpha$, where $S\propto \nu^{\alpha}$.

\section{Observations, Calibration and Imaging}\label{Observations, Calibration, and Imaging}

\subsection{Target Fields}\label{sec:target}

\begin{table*}
\scriptsize
 \centering
  \caption{Available radio data and observations in CDFS and ELAIS-S1. Listed is the wavelength, telescope and survey where applicable, area overlap with the data presented in this paper, sensitivity, angular resolution and reference.}\label{tab:surveys}\label{Tab:otherdata}
 \begin{tabular}{lllllll}
 \hline
 Field & Wavelength & Survey/ & Overlapping & Sensitivity & Resolution & Reference\\
  & & Instrument & Area (deg$^2$) & ($\mu$Jy/bm) & (\arcsec) & \\
 \hline
CDFS & $15\,$mm ($20\,$GHz) & ATCA & 2.50 & $300$ & $29.1\times 21.9$ & \citet{Franzen2014}\\   
 &  $6\,$cm ($5.5\,$GHz) & ATLAS/ATCA & 0.25 & $12$ & $4.0\times 2.0$ & \citet{Huynh2012b}\\
 &  $13\,$cm ($2.3\,$GHz) & ATLAS/ATCA & 3.57 & $80$ & $57.2 \times 22.7$ & \citet{Zinn2012}\\
 &  $21\,$cm ($1.4\,$GHz) & VLA & $0.33$ & $7.4$ & $2.8\times 1.6$ & \citet{Miller2013}\\
 &  $21\,$cm ($1.4\,$GHz) & ATLAS/ATCA & 3.57 & $40$ & $11\times 5$ & \citet{Norris2006}\\
 &  $21\,$cm ($1.4\,$GHz) & ATLAS/ATCA & 3.57 & $30$ & $12\times 6$ & \citet{Hales2014a}\\
 &  $21\,$cm ($1.4\,$GHz) & NVSS/VLA &3.57 & $450$ & 45 & \citet{Condon1998}\\
 &  $92\,$cm (325\,MHz)$^\ast$ & GMRT & 3.57 & $100$ & $7.1\times 10.8$ & \citet{Afonso2007} \\
 &  $1.25-3.75$\,m ($80-240$~MHz)$^\ast$ & GLEAM & 3.57 & 3000 & 150 & \citet{wayth2015} \\ 
 &  $1.5-3$\,m ($104-196$~MHz) & MWACS & 3.57 & 40000 & 180 & \citet{hurley-walker2014} \\ 
&   2\,m (150\,MHz) & TGSS/GMRT & 3.57 & $5-7$\,mJy & 20 & \citet{TGSS}\\
 &   4\,m (74\,MHz) & VLSS/VLA & 3.57 & 0.1\,Jy & 80 & \citet{Cohen2007}\\ 
  \hline
ELAIS-S1 &  $13\,$cm ($2.3\,$GHz) & ATLAS/ATCA & 2.70 & $70$ & $33.6 \times 19.9$ & \citet{Zinn2012}\\
 &   $21\,$cm ($1.4\,$GHz) & ATLAS/ATCA & 2.70 & $30$ & $10.3\times 7.2$ & \citet{Middelberg2008}\\
 &  $21\,$cm ($1.4\,$GHz) & ATLAS/ATCA & 2.70 & $30$ & $12\times 6$ & \citet{Hales2014a}\\
 &   $36\,$cm ($843\,$MHz) & MOST &   2.70 & $600$ & $62 \times 43$ & \citet{Randall2012}\\
 &   $50\,$cm ($610\,$MHz) & GMRT & 3.30 & 100 & $10 \times 5$ & \citet{Intema2014}\\
 &   $2\,$m ($150\,$MHz) & TGSS/GMRT & 2.70 & $5-7$\,mJy & 20 & \citet{TGSS}\\
 &  $1.25-3.75$\,m ($80-240$~MHz)$^\ast$ & GLEAM & 2.70 & 3000 & 150 & \citet{wayth2015} \\ 
 &   $1.5-3$\,m ($104-196$~MHz) & MWACS & 2.70 & 40000 & 180 & \citet{hurley-walker2014} \\
  \hline
 & $^\ast$ Data yet to be released.\\
\end{tabular}
\end{table*}

\begin{figure}
\includegraphics[scale=0.60]{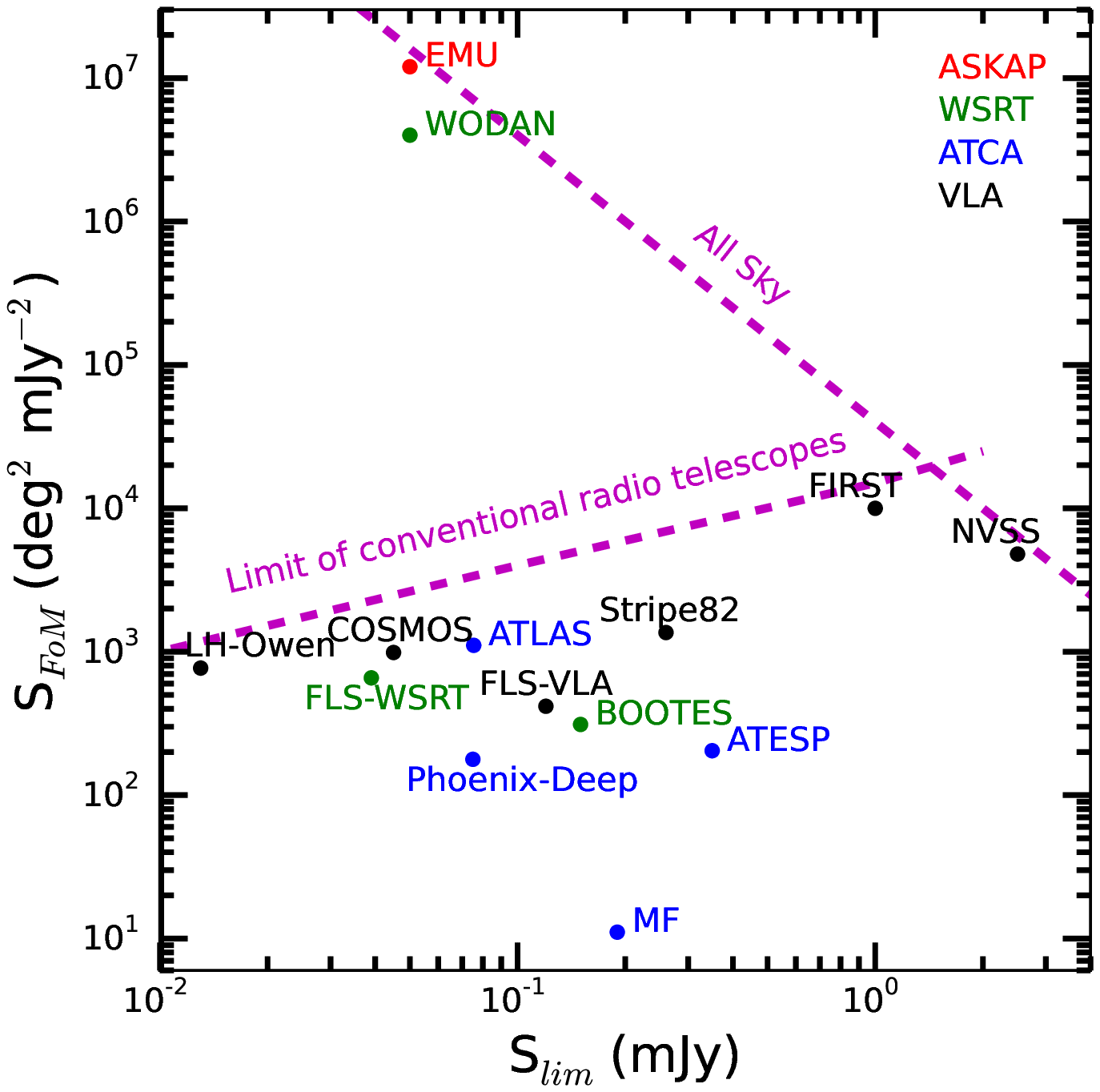}
\caption{Comparison of ATLAS with the following $1.4\,$GHz surveys: the Evolutionary Map of the Universe \citep[EMU;][]{Norris2011} survey, 
the Westerbork Observations of the Deep APERTIF Northern-Sky \citep[WODAN;][]{rottgering2010} survey, the NRAO VLA Sky Survey \citep[NVSS;][]{Condon1998}, 
the Faint Images of the Radio Sky Twenty Centimetres \citep[FIRST;][]{becker1995} survey, the Phoenix Deep Survey \citep{hopkins2003}, 
the Australia Telescope \textit{ESO} Slice Project \citep[ATESP;][]{Prandoni2000} survey and surveys of the Bootes field by \citet{devries2002},
the Marano Field by \citet{gruppioni1997}, the Lockman Hole by \citet{owen2008}, the COSMOS field by \citet{schinnerer2007}, 
Stripe 82 of the Sloan Digital Sky Survey by \citet{hodge2011} and the First-Look Survey region by \citet{morganti2004} and by
\citet{condon2003}. The telescope used to conduct each of the surveys is indicated in the top right. The horizontal axis is the $5\sigma$ detection limit of the surveys in mJy and the vertical axis is the $S_{\mathrm{FoM}}$ factor for surveys in deg$^2$ mJy$^{-2}$ (see section \ref{sec:target} for more detail). The two dashed lines indicate the All Sky limit and the approximate envelope of existing surveys, heavily dependent on available telescope time.}
\label{fig:survey}
\end{figure}

ATLAS covers two regions, each of $\sim 3~\mathrm{deg}^{2}$, surrounding CDFS ($\alpha = 03^{\rm h} 30^{\rm m} 16.3^{\rm s}$, $\delta = -28^{\rm o} 05' 12.4''$) and ELAIS-S1 ($\alpha = 00^{\rm h} 33^{\rm m} 50.8^{\rm s}$, $\delta = -43^{\rm o} 44' 57.4''$). These two fields have previously been observed at $1.4\,$GHz as part of the ATLAS project and are described by \citet{Hales2014a}.  Both of these fields were originally targeted for radio observations as they overlap with SWIRE which includes infrared and optical data for the majority of the radio objects.  Table \ref{tab:surveys} lists the available radio observations ($74\,$MHz -- $20\,$GHz) overlapping the two ATLAS fields.  

To compare ATLAS with other existing $1.4\,$GHz surveys shown in Fig.~\ref{fig:survey}, we compile a figure of merit for surveys, $S_{\mathrm{FoM}}$. The goal of a survey is to maximise the area observed and minimise the noise in the image. This is limited by the fact that it takes $T^{2}$ times longer to decrease the thermal noise by a factor of $T$ and $T$ times longer to increase the observing area by a factor of $T$. Therefore, $S_{\mathrm{FoM}}$ is of the form:
\begin{equation}
S_{\mathrm{FoM}} = \frac{\Omega}{(S_{\rm lim})^2} \, ,
\end{equation}
where $\Omega$ is the survey area in square degrees and $S_{\rm lim}$ is the $5\sigma$ detection limit of the survey in mJy. \cite{bunton2010} used a similar metric to
quantify the survey speed for chequerboard phased array feeds. Of the current 1.4~GHz surveys, FIRST has the largest $S_{\mathrm{FoM}}$ ($1\times10^{4}\,$deg$^2$ mJy$^{-2}$). ATLAS has a similar $S_{\mathrm{FoM}}$ (1108\,deg$^2$ mJy$^{-2}$) to LH-Owen, COSMOS and Stripe82, thereby providing among the deepest and widest coverage of radio objects at 1.4~GHz. 

We used the Australia Telescope Compact Array \citep[ATCA;][]{Frater1992} at $1.4\,$GHz to observe the two ATLAS fields in all four Stokes parameters ($XX, YY, XY, YX$). During our observing campaign, the ATCA was being upgraded with the Compact Array Broadband Backend \citep[CABB;][]{Wilson2011} providing a larger instantaneous bandwidth coupled with increased sensitivity of continuum and spectral line observations. At the time of our ATLAS observations, the CABB band provided a $500\,$MHz bandwidth covering $1.3-1.8\,$GHz, split into $1\,$MHz channels. 

ATCA project C1967 was allocated 1000 hours distributed over 78 days between 2009 June and 2010 June to extend ATLAS. The observations were spread over the four 6\,km array configurations to maximise {\it uv} coverage as listed in Table~\ref{Tab:obsinfo}. The primary flux and bandpass calibrator PKS 1934-638 \citep[$14.95\,$Jy at $1.380\,$GHz;][]{Reynolds1994} was visited at the beginning of each observing run. Two different secondary phase calibrators, PKS 0022-433 for ELAIS-S1 and PKS 0400-319 for CDFS, were observed every 15 minutes throughout the observations to determine the antenna complex gains and polarization leakage correction.  We observed the two fields in the standard ATCA mosaic mode with 28 pointings in CDFS and 20 pointings in ELAIS-S1.  The 48 pointing centres are listed in Table~\ref{Tab:pntinfo}.

\subsection{Calibration}

We calibrated and edited the {\it uv} data using {\tt MIRIAD}\footnote{http://www.atnf.csiro.au/computing/software/miriad/} \citep{Sault1995}.  The standard {\tt MIRIAD} calibration techniques are optimised for the original ATCA correlator system ($2\times128\,$MHz bandwidths) and had to be expanded to calibrate the new CABB-enabled bandwidth.  We calibrated each day of observations separately following the method outlined below.

We removed channels affected by self-interference due to the 640-MHz clock harmonics and by known radio frequency interference with {\tt atlod} in {\tt MIRIAD} using the options {\tt birdie} and {\tt rfiflag}.  We then restricted the frequencies to the known range of good bandpass response using {\tt uvaver} to include only data between 1.3 and 1.8~GHz.  Calibration of the bandpass was completed using the standard {\tt MIRIAD} procedure {\tt mfcal} and we applied the calibration to the secondary calibrator.  Automatic flagging of the calibrators was completed using {\tt mirflag}, setting {\tt int=600}.  This task implements the automatic flagging routine {\tt pieflag} developed by \citet{Middelberg2006}.  We also manually flagged the calibrators using {\tt uvflag}.

The large bandwidth poses an issue with frequency dependent calibration.  The standard {\tt MIRIAD} calibration procedure was altered to deal with the large bandwidth when calibrating the antenna gains and phases and the instrumental polarization.  The task {\tt gpcal} has been updated to allow for calibration over smaller frequency bands instead of over the full CABB band.  For our data, we calibrated over $128\,$MHz sections across the band using the option {\tt nbins} in {\tt gpcal}.  Once the calibration was complete, we copied the solutions to the targets using {\tt gpcopy}.  The target pointings were then flagged using {\tt mirflag} with {\tt int=1200} and further manually flagged with {\tt uvflag}.  The mean integration time per pointing after flagging is $\approx 15$ hours in CDFS and $\approx 9$ hours in ELAIS-S1.

\subsection{Imaging}\label{sec:Imaging}

\subsubsection{Single Pointing Image Processing}

We combined all data for each pointing into one file using {\tt uvcat} and then each pointing was imaged separately.  Since the observations cover a large fractional bandwidth, multifrequency synthesis was used.  

\citet{Sault1999} describe the process of multi-frequency synthesis and how to successfully {\tt CLEAN} an image.  When creating an image over a wide bandwidth, two dirty beam images must be used, as the spectrum of a radio sources is given by
\begin{equation}
I(\nu) = I(\nu_{\rm 0}) + \alpha I(\nu_{\rm0})\frac{\nu - \nu_{\rm 0}}{\nu_{\rm 0}} \ ,
\end{equation}
where $I$ is the flux density at frequency $\nu$, $\nu_{\rm 0}$ is the reference frequency and $\alpha$ is the spectral index \citep{Sault1999}.  The two dirty beams are the {\it synthesised dirty beam} and the {\it spectral dirty beam}.  The dirty image is then represented by
\begin{equation}
I_{\rm D}(l,m) = I(l,m)*B_{\rm 0}(l,m) + (\alpha (l,m)I(l,m))*B_{\rm 1}(l,m) \ ,
\end{equation}
where $B_{\rm 0}$ is the synthesised dirty beam, $B_{\rm 1}$ is the spectral dirty beam and $(l,m)$ are the directional cosines \citep{Sault1999}.  The {\tt MIRIAD} task {\tt invert} provides the option of creating the two dirty beams. These two dirty beams are used in the multi-frequency {\tt CLEAN} task, {\tt mfclean}, to create the {\tt CLEAN} component $I$ and $\alpha I$ maps. 

Each pointing was {\tt CLEAN}ed and self-calibrated separately prior to mosaicking. A model for self-calibration was produced by imaging the full band with uniform weighting and a cell size of 1~arcsec. The band was then split up into four 125~MHz subbands and self-calibration was applied to each subband separately in order to allow for variations in the antenna gains with frequency. Three iterations of phase self-calibration were applied, progressively increasing the number of {\tt CLEAN} components used to model the sky emission, and hence the total amount of flux included in the model. 

The final self-calibrated $uv$ data were then divided into two subbands of $250\,$MHz, imaged separately and adjusted to the reference pointing. Splitting up the data into two subbands serves to improve the accuracy of the primary beam correction. In order to obtain nearly identical synthesised beams for all pointings, individual `robust' weighting factors \citep{Briggs1995} were assigned. Pointings in the lower subband typically have a robust weighting factor in the range 0.0--0.5 and in the higher subband 0.4--1.0. 

Using robust weighting factors in these ranges was found to minimise the rms noise in the final mosaic. Using higher robust weighting factors -- a robust weighting factor of 2 or more corresponds very closely to natural weighting -- was found to degrade the image sensitivity, despite the decrease in the theoretical noise level. This probably arises, in large part, from the degradation of the sidelobe levels and beam-shape, which renders the task of {\tt CLEAN}ing the image and removing artifacts harder. 

The larger beam size obtained when using natural weighting also increases the confusion noise \citep{Condon2012}. The final beam size in ELAIS-S1 is 12.2 by 7.6~arcsec and in CDFS is 16.3 by 6.8~arcsec (see Section~\ref{mosaics}). At a frequency of 1.4~GHz and a resolution of 9.6~arcsec (the geometric mean of the major and minor beam axes in ELAIS-S1), the confusion noise $\sigma_{\mathrm{c}} \approx 3.8~\mu$Jy beam$^{-1}$. $\sigma_{\mathrm{c}}$ is predicted to vary as $\theta^{10/3}$, where $\theta$ is the beam size. Using a robust weighting factor of 2 for all pointings in ELAIS-S1 results in a final beam size of 17.9 by 11.8~arcsec, for which $\sigma_{\mathrm{c}} \approx 14.9~\mu$Jy beam$^{-1}$.  

When imaging the self-calibrated $uv$ data, the cell size was set to 1.5~arcsec. We {\tt CLEAN}ed each pointing down to 5$\sigma$. A 2D elliptical Gaussian fitted to the central region of the dirty beam was used as the restoring beam. The {\tt mfs} option in {\tt restor} was used to write a second plane in the output image containing the $\alpha I$ model convolved with the Gaussian beam. The $\alpha I$ plane was subsequently used to perform wideband primary beam correction in {\tt linmos} (see Section~\ref{mosaics}). The pointing reference for ELAIS-S1 is $\alpha = 00^{\rm h} 33^{\rm m} 50.8^{\rm s}$, $\delta = -43^{\rm o} 44' 57.4''$ and for CDFS is $\alpha = 03^{\rm h} 30^{\rm m} 16.3^{\rm s}$, $\delta = -28^{\rm o} 05' 12.4''$.

\subsubsection{Image Artifacts}\label{artifacts}

\begin{figure}
 \begin{center}
  \includegraphics[scale=0.35, angle=270]{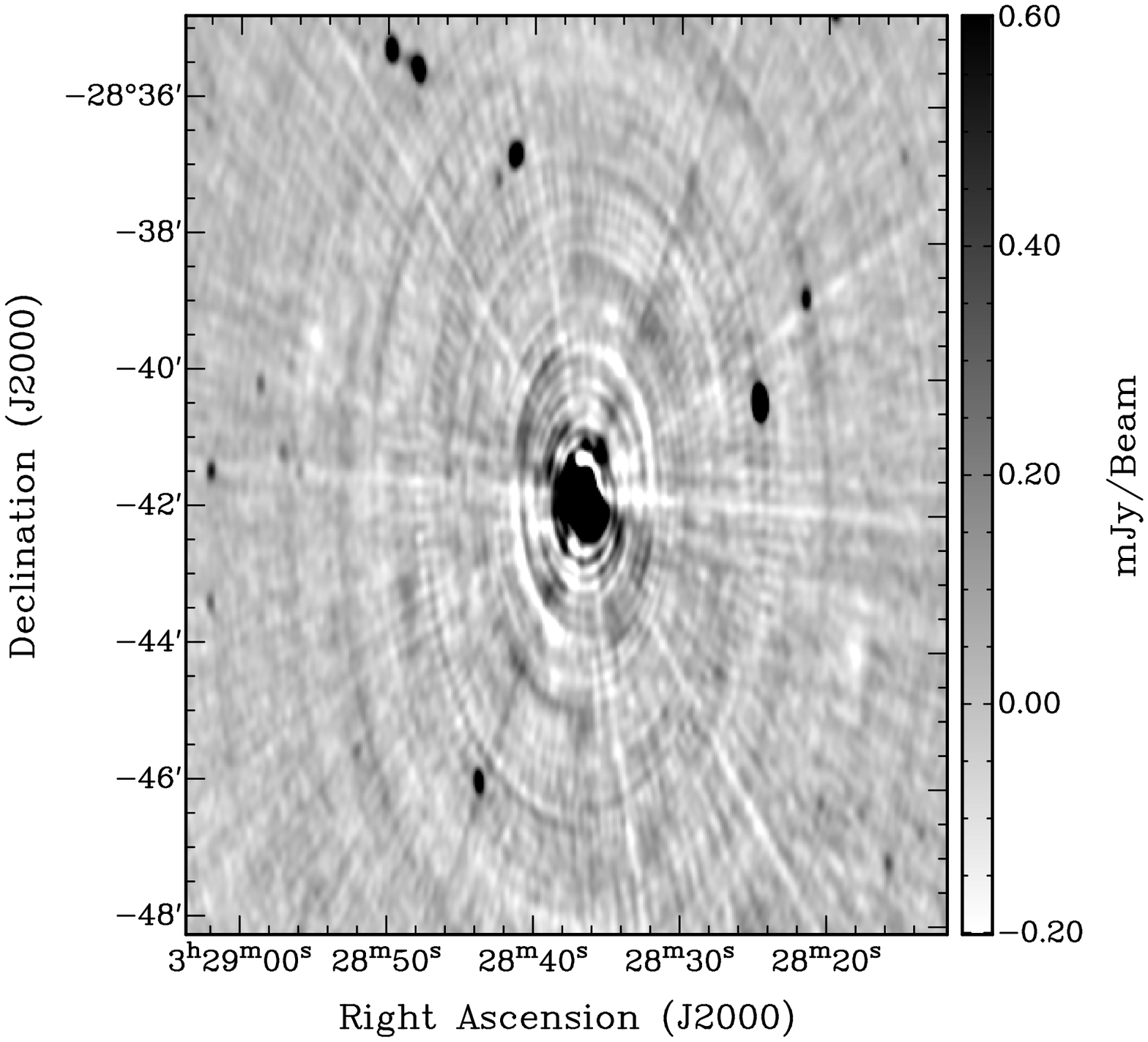}
  \includegraphics[scale=0.35, angle=270]{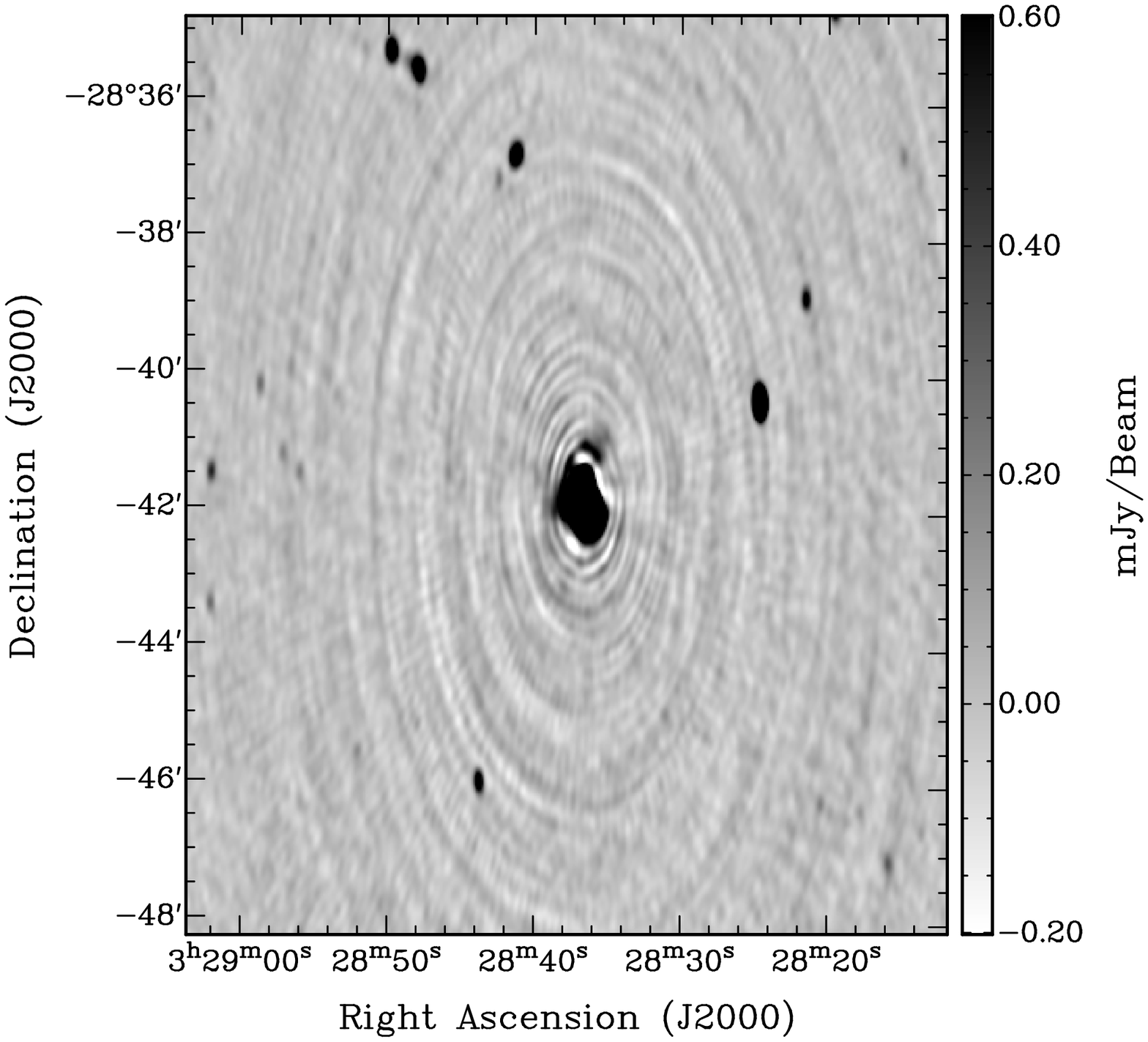}
  \caption{Map of a 1.4~Jy source (NVSS J032836-284147) in CDFS, before (top) and after (bottom) amplitude self-calibration.}
  \label{fig:amplitude_selfcal}
 \end{center}
\end{figure}

Radial spokes are present around bright sources in both fields. We found that these can be effectively removed using amplitude self-calibration. Fig.~\ref{fig:amplitude_selfcal} shows the improvement in dynamic range after amplitude self-calibration around a 1.4~Jy source \citep[NVSS J032836-284147;][]{Condon1998} in CDFS; the dynamic range (as measured by dividing the peak flux density of the source by the rms in the vicinity of the source) is $\approx 70$ per cent higher after amplitude self-calibration. 

We tried to apply amplitude self-calibration to the ATLAS pointings using components with flux density greater than $5 \sigma$ to model the sky brightness distribution. This caused the flux densities of sources below approximately 1~mJy to be biased low. The magnitude of the bias was found to increase for sources with decreasing flux density below 1~mJy, reaching $\sim 10$ per cent close to the $5 \sigma$ detection limit. We suspect that a considerable amount of flux was missing in the model due to the flat slope of the counts at sub-mJy levels, resulting in the observed bias in the source flux densities. \cite{wieringa1992} investigated self-calibration methods in use at radio synthesis arrays and found that self-calibration can bias the gains if a significant amount of flux is missing in the model, particularly in cases where the number of antennas is low. Any sources not fully contained in the model tend to get absorbed into the gains and are then reduced in amplitude in the image after self-calibration. Amplitude self-calibration was therefore not used when producing the final images.

Mild artifacts in the form of concentric rings are also visible around the strong source shown in Fig.~\ref{fig:amplitude_selfcal}. 
These limitations are related to difficulties in modelling and calibrating time-dependent effects simultaneously using standard
self-calibration techniques, particularly 
where the source is bright and partially (or fully) resolved. This is primarily due to the east-west nature of ATCA array 
configurations (i.e. at any point in time only a slice in the $uv$ plane is sampled). \cite{lenc2009} found that, in the pointing 
closest to the strong source, these artifacts could be removed by modelling the source in the $uv$ plane with a combination of 
Gaussians and point-like components, using \textsc{Difmap} \citep{shepherd1997}. We did not apply this technique here because it 
was difficult to automate and it could not handle wideband data.

\subsubsection{Addition of Previous ATLAS Data}

Previous ATLAS DR2 observations (ATCA Project IDs C1035 and C1241) from \cite{Hales2014a} were combined with our CABB observations to maximise the sensitivity.
Information on these observations including observing dates and ATCA array configurations is given in Table \ref{Tab:obsinfo}.
The data were combined in the image plane for reasons related to bandwidth smearing, and described in detail in Section~\ref{sec:bws}.  These data consist of two $128\,$MHz bands centred at 1344 and $1432\,$MHz each containing 16 channels of $8\,$MHz in size. Both bands were imaged together using a similar procedure to the CABB data.  The robust weighting factors for the pointings were chosen to yield a similar beam size to that of the CABB data. Pointings typically have a robust weighting factor in the range $-0.1$ -- $+0.1$.

Addition of the pre-CABB data results in an improvement in sensitivity of $\approx 20$ per cent and also improves the dynamic range thanks to the increased $uv$ coverage.

\subsection{Mosaics}\label{mosaics}

\begin{figure*}
\begin{center}
\includegraphics[scale=0.8,angle=270, trim=2.5cm 0.5cm 1cm 0cm]{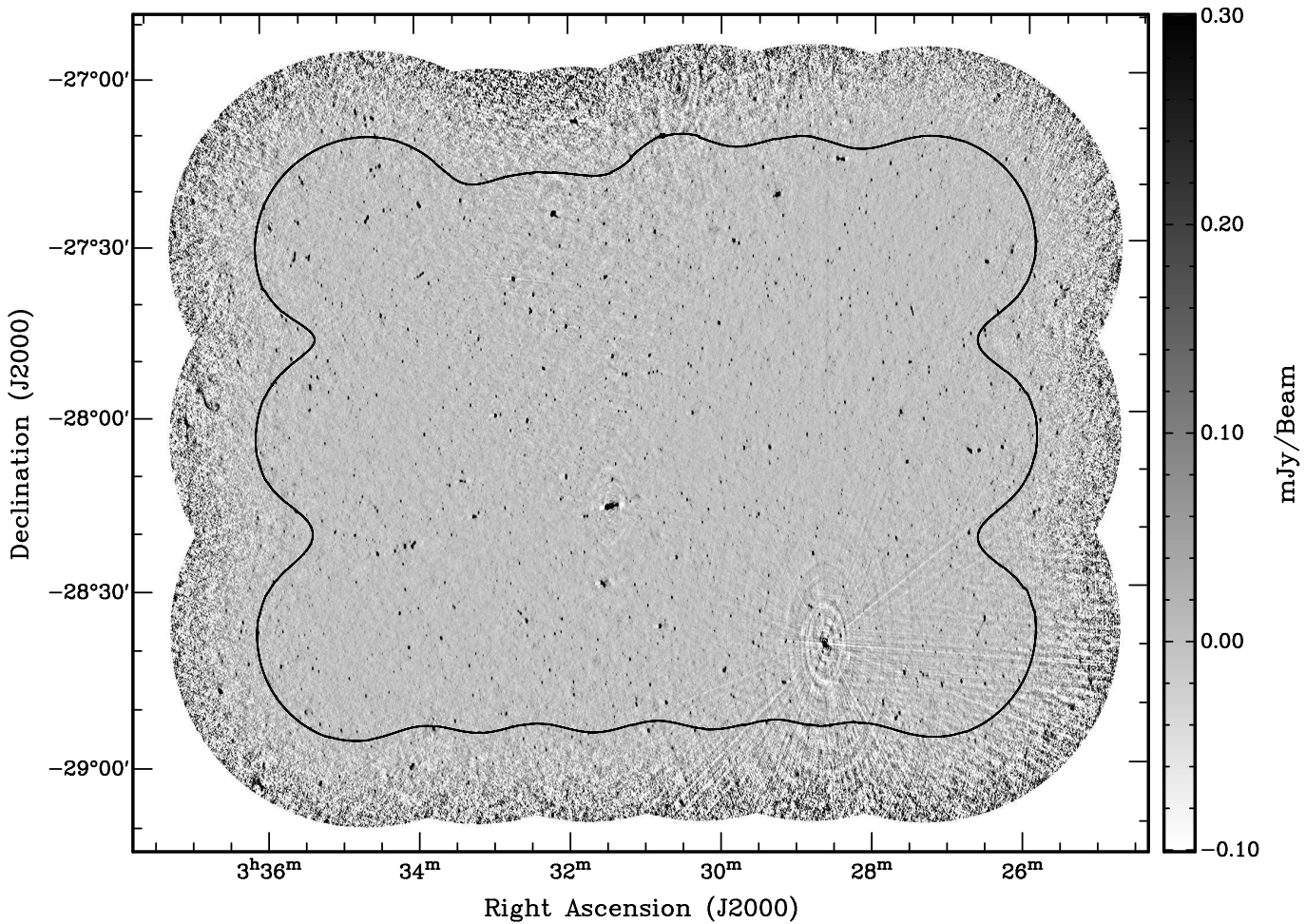}
\caption{The ATLAS CDFS total intensity mosaic with the linear greyscale set to the range $-100$ to $+300\,\mu$Jy beam$^{-1}$. The image projection used is North-celestial-pole \citep[NCP;][]{greisen1983}, a projection onto a plane perpendicular to the North Celestial Pole, which is a special case of the orthographic (SIN) projection, often used for east-west radio interferometers. The solid black contour indicates the component catalogue boundary (3.6~$\mathrm{deg}^{2}$) of the mosaic defined by: (1) local rms noise $\le 100\,\mu$Jy beam$^{-1}$; (2) sensitivity loss due to bandwidth smearing $< 20$ per cent; and (3) mosaicked primary beam response $\ge 40$ per cent. The pattern of pointings on the sky is identical to that shown in \citet{Norris2006}.}
\label{fig:CDFSmos}
\end{center}
\end{figure*}

\begin{figure*}
\begin{center}
\includegraphics[scale=0.75,angle=270, trim=2cm 0cm 1.2cm 0cm]{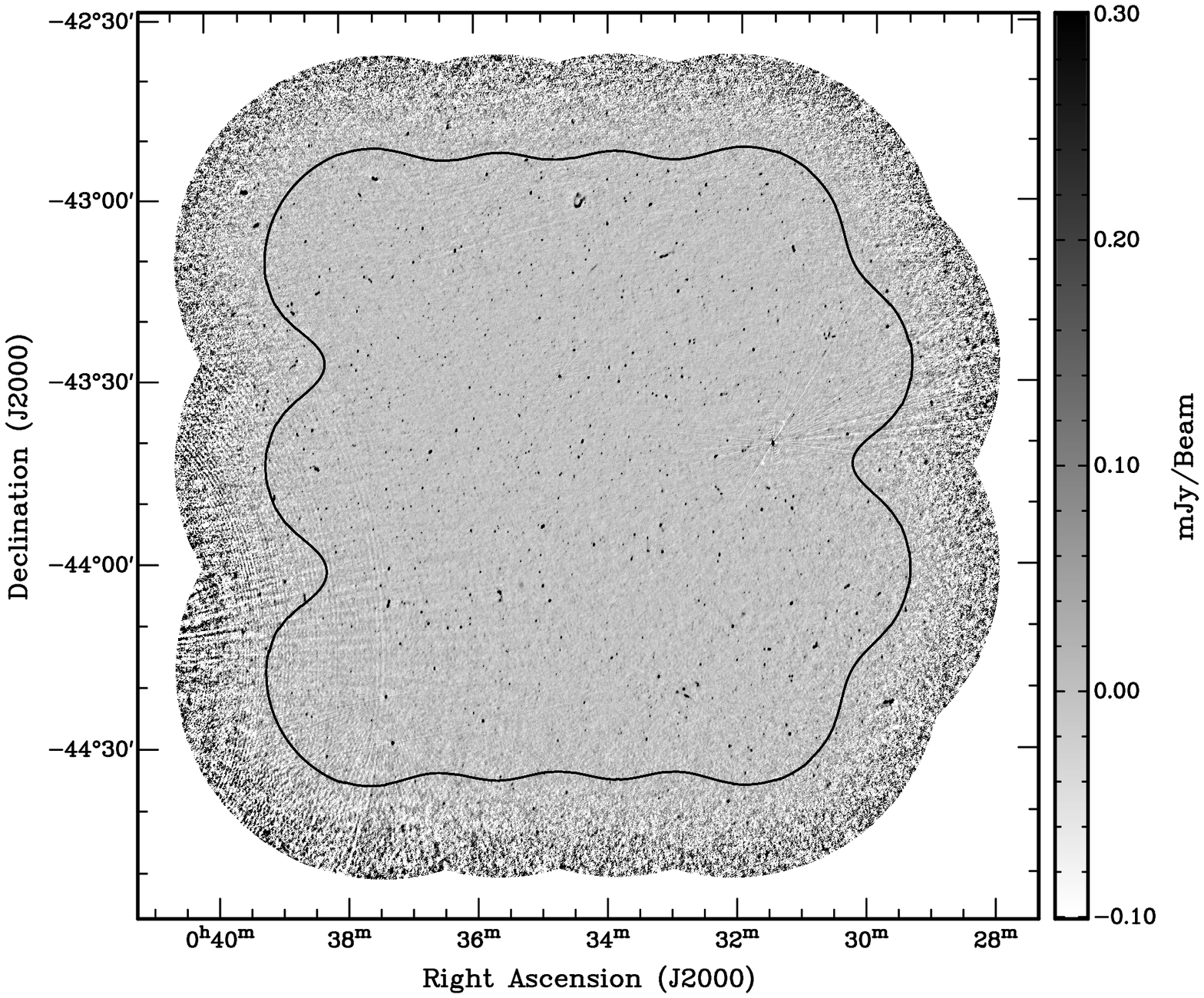}
\caption{The ATLAS ELAIS-S1 total intensity mosaic with the linear greyscale set to the range $-100$ to $+300\,\mu$Jy beam$^{-1}$. The image projection used is NCP. The solid black contour indicates the component catalogue boundary (2.7~$\mathrm{deg}^{2}$) of the mosaic defined by: (1) rms noise $\le 100\,\mu$Jy beam$^{-1}$; (2) sensitivity loss due to bandwidth smearing $< 20$ per cent; and (3) mosaicked primary beam response $\ge 40$ per cent. The pattern of pointings on the sky is identical to that shown in \citet{Middelberg2008}.}
\label{fig:ELAISmos}
\end{center}
\end{figure*}

Each pointing was convolved with a Gaussian to obtain an identical synthesised beam across each of the two fields separately. CDFS has a beam size of 16.3 by 6.8~arcsec, with position angle $2\degr$, and ELAIS-S1 a beam size of 12.2 by 7.6~arcsec, with position angle $-11\degr$. A source-free estimate of the noise in each pointing image was obtained as follows: an initial estimate of the noise was obtained by taking the rms within the primary beam full-width at half-maximum (FWHM). To avoid the noise estimate from being affected by real source emission, all pixels outside the range $\pm 3\sigma$ were then flagged and the rms was re-evaluated. This process was repeated a number of times until the noise was found to decrease by less than 10 per cent after removing pixel outliers.

Finally, using {\tt linmos}, the pointings for each field were corrected for the primary beam and mosaicked together, weighting them by their respective rms noise values; Gaussian primary beam fits for the new 16~cm CABB receivers across the entire frequency range (1.1 to 3.1~GHz)\footnote{http://www.narrabri.atnf.csiro.au/people/ste616/beamshapes/beamshape\_16cm.html} were used and the primary beam response was averaged over the subband as described below. Fig.~\ref{fig:CDFSmos} shows the resulting CDFS total intensity (Stokes $I$) image and Fig.~\ref{fig:ELAISmos} shows the resulting ELAIS-S1 total intensity image.

Due to the wideband nature of the observations there is a discrepancy between the integrated flux density over the band (as returned by {\tt MIRIAD}) and the monochromatic flux density at the central frequency for any source with a non-zero spectral index. This is essentially due to most sources being best described by a power-law slope across the band rather than a simple linear slope. The integrated flux density for a source with a power-law rather than a linear slope across the band is always going to exceed the monochromatic flux density at the central frequency. For small $\frac{\Delta\nu}{\nu_0}$ this effect is small, but increases for wider bands and for a source with an increasingly non-flat spectrum. In our case, for $\alpha=-0.75$, typical of radio sources whose emission is dominated by optically thin synchrotron radiation, a source would have its monochromatic flux density over-estimated by less than 0.5 per cent in each subband, well within the absolute calibration errors.

However, this issue gets more complicated off-axis as the attenuation of the primary beam drops off more rapidly at higher frequencies effectively reducing the sensitivity at higher frequencies. For narrow-band observations with a well known primary beam pattern (which varies with frequency), using the primary beam correction factor at the central frequency is sufficient. For wide-band observations the primary beam correction at this frequency is not truly representative. 

To improve the accuracy of the primary beam correction, we used the implementation of wideband primary beam correction in {\tt linmos} (option {\tt bw}): 
the image and primary beam were evaluated at 10 frequencies, the image using the $\alpha I$ plane and the beam using the beam fits across the frequency range. 
All these images and beams were then used in the standard mosaic equation \citep[see e.g.][]{sault1996} to produce the final mosaicked image.

\section{Correction of Final Mosaics}\label{Correction of Final Mosaics}

\subsection{CLEAN Bias}

\begin{table}
 \centering
  \caption{Number of simulated sources added to the {\it uv} data of one ELAIS-S1 pointing at a given SNR to analyse the CLEAN bias.}\label{cleanbias}
 \begin{tabular}{r r}
 \hline
 $N$ & SNR \\
 \hline
 40 & 5 \\
 15 & 6 \\
 15 & 7 \\
  15 & 8 \\
 15 & 9 \\
 10 & 10 \\
10 & 12\\
 5 & 16\\
 3 & 20\\
 2 & 30\\
 1 & 50\\
1 & 100\\
  \hline
\end{tabular}
\end{table}

\begin{figure}
\begin{center}
\includegraphics[scale=0.35,angle=270]{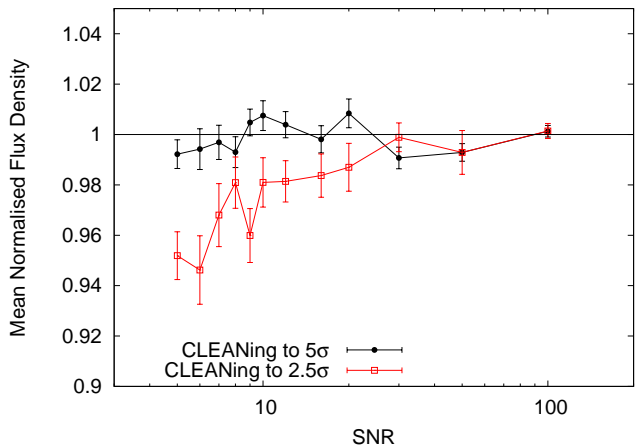}
\caption{The mean normalised flux density of simulated sources as a function of the SNR.  The lines illustrate the {\tt CLEAN} bias when {\tt CLEAN}ing to $5\sigma$ (black line and circles) and $2.5\sigma$ (red line and squares).}
\label{fig:clean_bias}
\end{center}
\end{figure}

{\tt CLEAN} bias \citep{Condon1998} is an effect in deconvolution which redistributes flux from point sources to noise peaks in the image, thereby resulting in a systematic underestimation of the flux densities of real sources. As the amount of flux which is taken away from a real source is independent of its flux density, the fractional error this causes is largest for faint sources. The magnitude of the effect will depend on the {\it uv} coverage and to what flux density level the images are {\tt CLEAN}ed.

In order to analyse the {\tt CLEAN} bias on the flux density of sources in our ATLAS mosaics, we followed a similar procedure to that outlined in \citet{Middelberg2008}. We added 132 point sources at random positions to the $uv$ data of one ELAIS-S1 pointing, except that a simulated source could not lie closer than 1~arcmin from a real source ($>5\sigma$) and any other simulated source. The numbers of sources added with different signal-to-noise ratios (SNRs) are listed in Table \ref{cleanbias}.  We then imaged the data in the same way as the final image (see Section~\ref{sec:Imaging}), extracted the flux density of each simulated source and divided by the injected flux density. This simulation was repeated 30 times to improve the statistical sampling.

The mean normalised flux density of the simulated sources as a function of SNR is shown in Fig.~\ref{fig:clean_bias}. The flux densities of the simulated sources were measured by extracting the map values at the simulated source positions, interpolated between pixels. This was done by calculating the map values on a successively finer grid (up to 128 times finer), by repeated convolution with a Gaussian-graded sinc function \citep{Rees1990}. At low SNR, the thermal noise will introduce a slight shift in the position of a point source, resulting in an increased likelihood for the peak to lie on top of a positive noise fluctuation. This, in turn, causes the peak flux density of a point source to be biased slightly high; this is known as the peak flux density bias. We chose to measure the flux densities of the simulated sources at their true positions, rather than measure their peak flux densities, to disentangle the peak flux density bias from the {\tt CLEAN} bias. The peak flux density bias is dealt with separately by the {\tt BLOBCAT} source finder in Section~\ref{sec:srcext}.

The black circles in Fig.~\ref{fig:clean_bias} show the {\tt CLEAN} bias when {\tt CLEAN}ing the images to 5$\sigma$, which typically corresponds to 5000 iterations (the final images were {\tt CLEAN}ed to this level). The simulations indicate that a source with an SNR of 5 would have its flux density underestimated by less than 1 per cent as a result of the {\tt CLEAN} bias, well within the calibration error of 5 per cent (see Section~\ref{sec:Calibration accuracy checks}). We therefore chose not to correct the sources' flux densities for this effect.

We repeated the simulations {\tt CLEAN}ing the images to 2.5$\sigma$ (this typically corresponds to 65000 iterations) to see how the {\tt CLEAN} bias is affected by the flux density level to which the images are {\tt CLEAN}ed. In this case, the {\tt CLEAN} bias is much more significant as shown by the blue squares in Fig.~\ref{fig:clean_bias}; the {\tt CLEAN} bias is $\sim$5 per cent for sources with an SNR of 5.
 
\subsection{Bandwidth Smearing}\label{sec:bws}

\begin{figure}
\begin{center}
\includegraphics[scale=0.45,angle=270, trim=1cm 3cm 0cm 0cm]{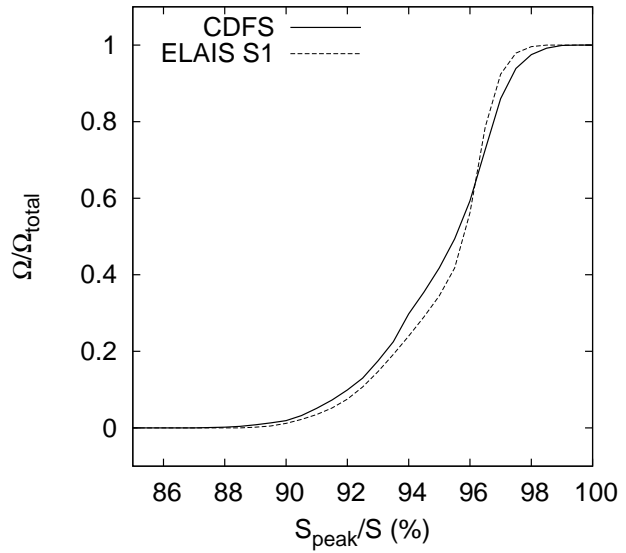}
\caption{The fraction of the CDFS (solid line) and ELAIS-S1 (dashed line) mosaics at or below a given bandwidth smearing level.  The total area of the mosaic ($\Omega_{\rm total}$) is defined where the rms noise level is less than $100\,\mu$Jy beam$^{-1}$.}
\label{fig:bws}
\end{center}
\end{figure}

Bandwidth smearing (chromatic aberration) is characterised by a smearing of the visibilities of a point source in the $uv$ plane due to the range of frequency being sampled over a given bandwidth. This results in a decrease in the peak flux density of the source \citep{Condon1998}.  The effect on sources in an image is to introduce smearing in the radial direction from the pointing centre, so although the peak flux density is reduced, the integrated flux density remains the same. Bandwidth smearing can cause the peak flux density to fall below the sensitivity threshold, thus rendering it unrecoverable. The magnitude of the effect will depend on the source distance from the pointing centre and the fractional channel bandwidth. The effect of bandwidth smearing is small for CABB data (with 1~MHz channels) but significant for pre-CABB data (with 8~MHz channels) and is the primary cause of the amplitude calibration discrepancy reported by \cite{Norris2006}.

In order to analyse the bandwidth smearing in our ATLAS mosaics, we followed a similar procedure to that outlined in \citet{Hales2014a}.  We modelled the amount of bandwidth smearing to produce simulated images from \citeauthor{Hales2014a}:
\begin{equation}
\frac{S_{\rm peak}}{S} = \left[1 + \frac{2 {\rm ln}2}{3}\left(\frac{\Delta\nu_{\rm eff}}{\nu}\frac{d}{B_{\rm proj(\zeta)}}\right)\right]^{-\frac{1}{2}} \ ,
\end{equation}
where $S_{\rm peak}$ is the peak flux density of the source, $S$ the integrated flux density of the source, $\Delta\nu_{\rm eff}$ the effective channel bandwidth, $\nu$ the central frequency, $d$ the radial distance from the phase centre and $B_{\rm proj}(\zeta)$ the projected beam FWHM for an elliptical beam.  The projected beam FWHM is given by \citeauthor{Hales2014a}:
\begin{equation}
B_{\rm proj}(\zeta) = \frac{B_{\rm maj}B_{\rm min}}{\sqrt{[B_{\rm maj}{\rm sin}(\zeta -\psi)]^2 + [B_{\rm min}{\rm cos}(\zeta -\psi)]^2}} \ ,
\end{equation}
where $\zeta$ is the position angle (East of North) of a source with respect to the phase centre, $B_{\rm maj}$ and $B_{\rm min}$ are the major and minor axes FWHM of the elliptical beam respectively, and $\psi$ is the beam position angle measured East of North.  

All of these simulated images were mosaicked together following the method outlined in Section~\ref{mosaics}. The effect of bandwidth smearing on the CDFS and ELAIS-S1 mosaics is shown in Fig.~\ref{fig:bws}.  The bandwidth smearing for both mosaics is less than 10 per cent over the great majority of the observational area. Sources' peak flux densities were corrected for bandwidth smearing, using our maps quantifying the sensitivity loss due to bandwidth smearing, as described in Section~\ref{sec:srcext}.

\section{1.4~GH\lowercase{z} Component Catalogue}\label{Source Component Catalog}
\subsection{Background Noise Map}\label{sec:noise}

\begin{figure}
\begin{center}
\includegraphics[scale=0.45,angle=270, trim=1cm 3cm 0cm 0cm]{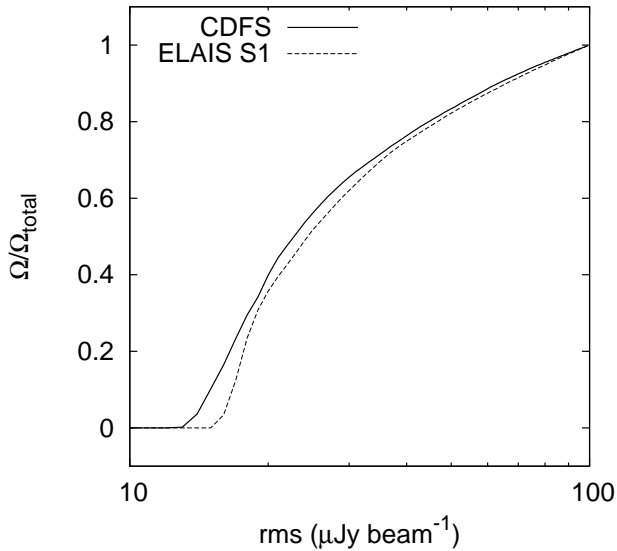} 
\caption{The fraction of the CDFS (solid line) and ELAIS-S1 (dashed line) mosaics at or below a given rms noise level.  The total area of the mosaic ($\Omega_{\rm total}$) is defined where the rms noise level is less than $100\,\mu$Jy beam$^{-1}$.}
\label{fig:noise}
\end{center}
\end{figure}

In order to extract sources from the mosaics we require an estimate of the background noise level.  The background rms noise of each mosaic was calculated for each pixel following the method described by \citet{Franzen2014}.  For each pixel, the noise was taken as the rms inside a box of size $2 l +1$ centred on the pixel; we chose the half-width of the box, $l$, to be 20 times the beam for both mosaics to minimise the number of spurious detections \citep{Huynh2012}. However, in order to avoid the noise estimate from being affected by real source emission, points were clipped iteratively until convergence at $\pm 3 \sigma$ was reached. Fig.~\ref{fig:noise} shows the fraction of the ATLAS sky at or below a given rms noise level, where the ATLAS sky is defined where the rms noise is less than $100\,\mu$Jy beam$^{-1}$. Table~\ref{tab:var_vs_alpha} gives statistics for the rms noise distribution across the ATLAS sky for the two fields.

\begin{table}
 \begin{center}
 \caption{Statistics for the rms noise distribution across the ATLAS sky (region where the rms noise is less than $100\,\mu$Jy beam$^{-1}$) for CDFS and ELAIS-S1.}
 \label{tab:var_vs_alpha}
 \begin{tabular}{@{} c c c }
 \hline
 & CDFS & ELAIS-S1 \\
 \hline
 Mean ($\mu$Jy beam$^{-1}$)  &  30  &   35  \\
 Median ($\mu$Jy beam$^{-1}$)  & 23   &  24 \\
 Mode ($\mu$Jy beam$^{-1}$)  &  14  & 17  \\
 \hline
 \end{tabular}
 \end{center}
\end{table}

\subsection{Source Extraction}\label{sec:srcext}
We limited the area of the two ATLAS fields for source extraction defined 
by the union of the following criteria: 
\begin{enumerate}
\item[(1)] rms noise $\le 100\,\mu$Jy beam$^{-1}$.
\item[(2)] sensitivity loss due to bandwidth smearing $< 20$ per cent.
\item[(3)] mosaicked primary beam response $\ge40$ per cent of the peak response.
\end{enumerate}
Fig.~\ref{fig:CDFSmos} shows the defined area covering 3.6~$\mathrm{deg}^{2}$ of the CDFS mosaic and Fig.~\ref{fig:ELAISmos} shows the defined area covering 2.7~$\mathrm{deg}^{2}$ of the ELAIS-S1 mosaic.  The resulting area is primarily defined by the primary beam response.

We extracted source components from the two ATLAS fields using {\tt BLOBCAT} \citep{Hales2012}, {\tt AEGEAN} \citep{Hancock2012}, {\tt pyBDSM}\footnote{http://dl.dropboxusercontent.com/u/1948170/html/index.html} and the AMI {\tt SOURCE\_FIND} software \citep{Franzen2011}. When using {\tt BLOBCAT}, {\tt AEGEAN} and {\tt pyBDSM}, we searched both mosaics down to an SNR $\ge 4$ in order to include all sources with fitted peak SNRs $\ge 5$. Sources with fitted peak SNRs $< 5$ were removed manually from the output component lists. When running {\tt SOURCE\_FIND} on the two mosaics, a detection threshold of 5$\sigma$ was used since the detection threshold is automatically lowered to ensure that all sources with fitted peak SNRs $\ge 5$ are included. We compared the source finders by locating the corresponding counterparts to within 20~arcsec.  The difference between the four source finders for our work is negligible.  However, we chose {\tt BLOBCAT} over the other source finders as {\tt BLOBCAT} takes into consideration bandwidth smearing and peak bias.  A more detailed analysis of the source finders is beyond the scope of this paper but is discussed by Hopkins et al. (2014), submitted.

Components that corresponded to artifacts around bright sources were removed from the {\tt BLOBCAT} catalogues for the two fields, resulting in a total of 2861 components for CDFS and 1964 for ELAIS-S1.

{\tt BLOBCAT} assumes that isolated components have Gaussian morphology in order to catalogue their properties. This assumption may not always be suitable for complex components. We identified complex blobs as having $N_{\mathrm{pix}} \geq 300$ and $R_{\mathrm{est}} \geq 1.4$, where $N_{\mathrm{pix}}$ is the number of flooded pixels comprising the blob and $R_{\mathrm{est}}$ is the size estimate of the blob, in units of the sky area covered by an unresolved Gaussian blob with the same peak flux density. Gaussian fitting was performed for each of these complex blobs, as recommended by \cite{Hales2012}, with the {\tt CASA} task {\tt imfit}. Initially, two Gaussians were fitted simultaneously and the residuals were examined. If the peak of the residuals was less than $5 \sigma$, then the fit was considered to be acceptable and the original {\tt BLOBCAT} catalogue entry was replaced by the {\tt imfit} entry for each individual Gaussian component. Otherwise, the number of Gaussians included in the fitting was increased by one and the residuals were re-evaluated. This process was repeated until the peak residual was found to be less than $5 \sigma$ or 8 Gaussians had been fitted.

Following this procedure, Gaussian fitting was performed for a total of 97 and 72 {\tt BLOBCAT} components identified as complex in CDFS and ELAIS-S1, respectively. We then merged together those Gaussian components which were separated by less than half the beam size. Our final component catalogue for CDFS contains 3034 components and that for ELAIS-S1 contains 2084 components.


\subsection{Deconvolution}\label{sec:res}

\begin{figure*}
 \begin{center}
  \includegraphics[scale=0.33,angle=270]{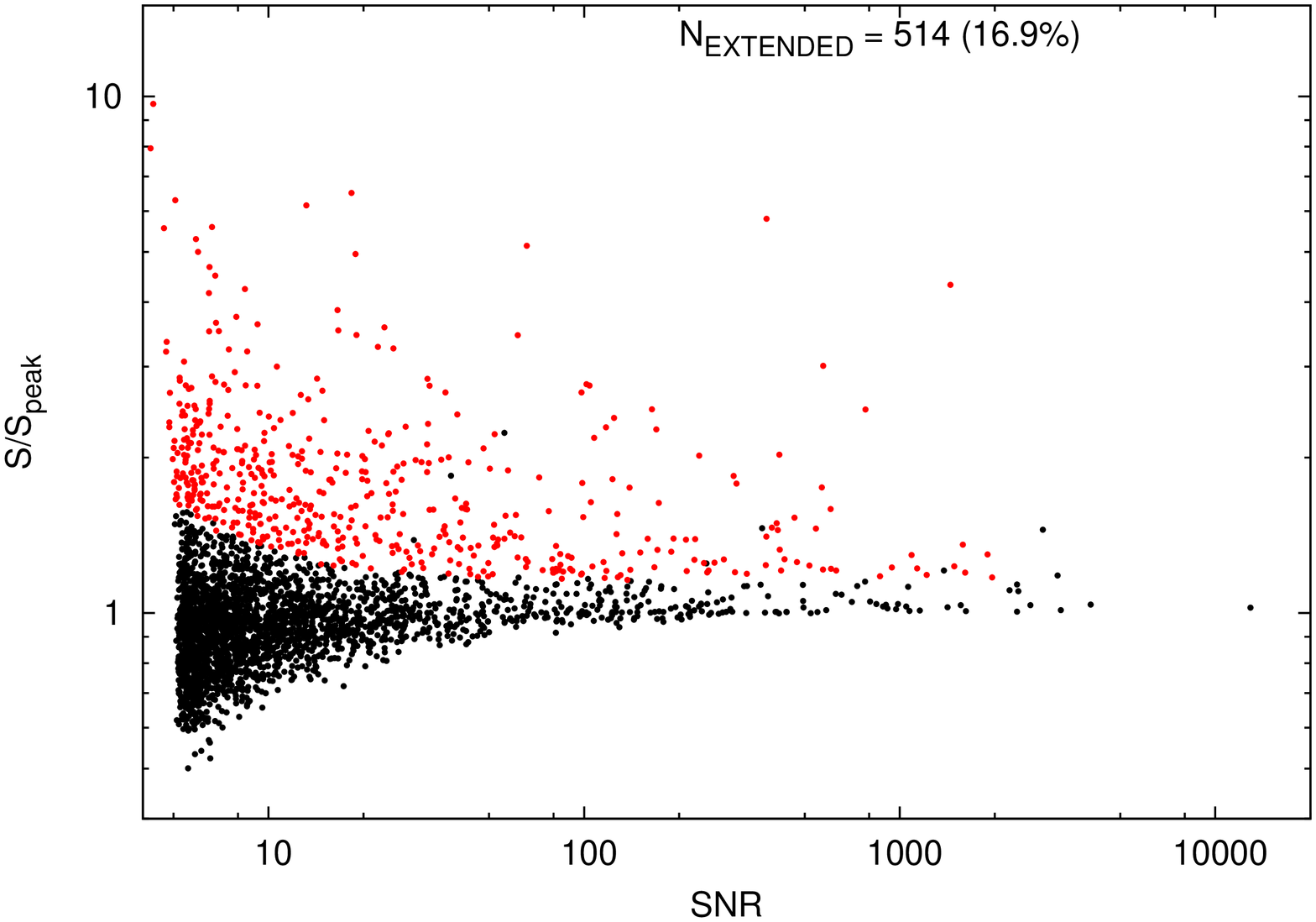}
  \includegraphics[scale=0.33,angle=270]{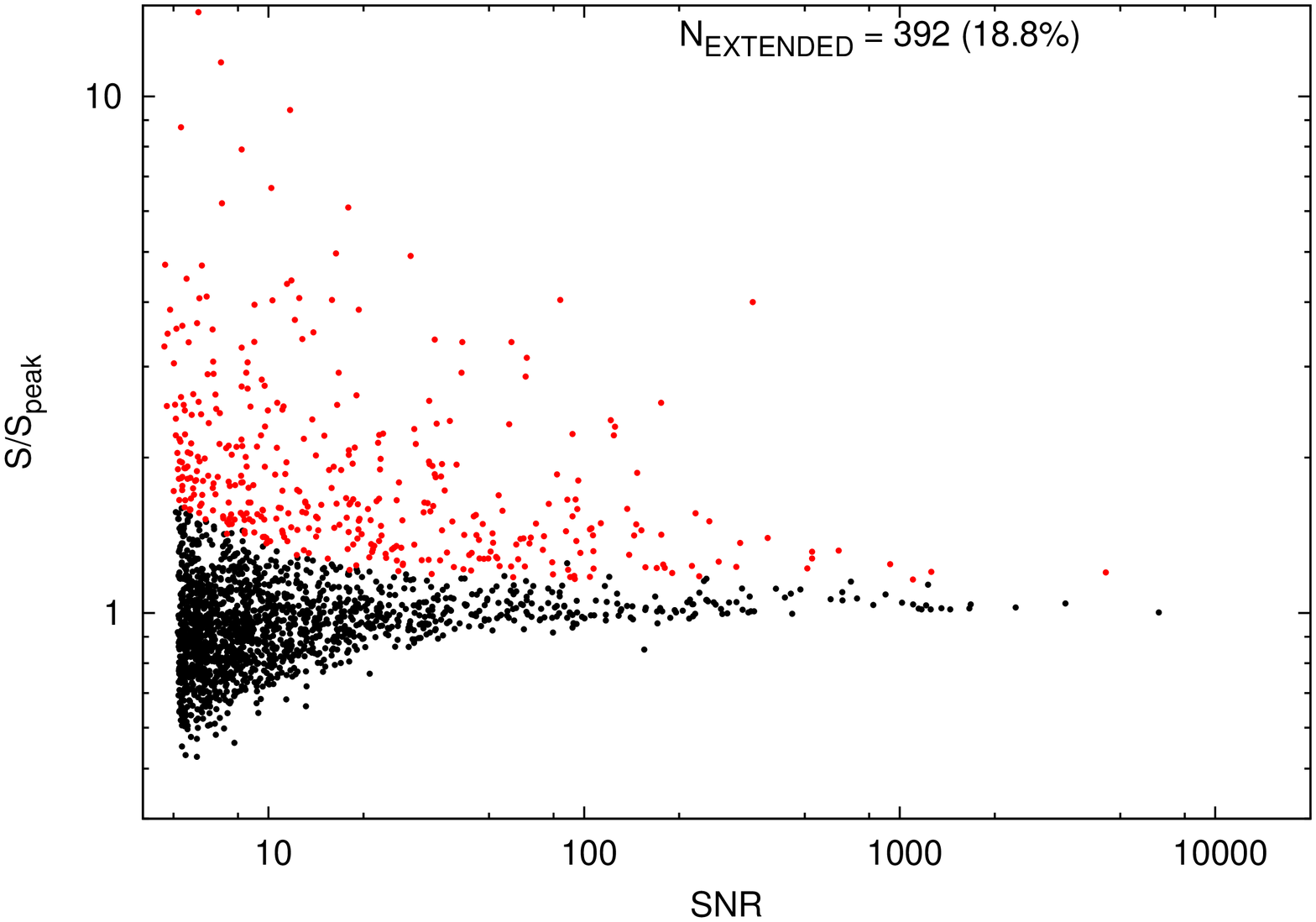}
  \caption{$S/S_{\mathrm{peak}}$ as a function of SNR for all components in CDFS (left) and ELAIS-S1 (right). The peak flux density values have been corrected for peak bias and bandwidth smearing as measured in {\tt BLOBCAT}. Components which are classified as point-like/extended are shown in black/red.}
  \label{fig:res}
 \end{center}
\end{figure*}

A measure of the degree to which a radio source is resolved is given by the ratio of the integrated flux density, $S$, to the peak flux density, $S_{\rm peak}$:
\begin{equation}\label{eqn:ratio}
\frac{S}{S_{\rm peak}} = \frac{\theta_{\rm min}\theta_{\rm max}}{B_{\rm maj}B_{\rm min}} \ ,
\end{equation}
where $\theta_{\rm min}$ and $\theta_{\rm max}$ are the source fitted FWHM axes.

In Fig.~\ref{fig:res}, we plot $S/S_{\mathrm{peak}}$ as a function of SNR for CDFS in the left panel and for ELAIS-S1 in the right panel. The peak flux density has been corrected for bandwidth smearing and peak bias. The distribution of components is skewed to higher flux density ratios at higher SNR, a result of extended components.  The occurrence of components with $S/S_{\mathrm{peak}} < 1$ are from noise fluctuations and calibration errors. For point sources, assuming that $\sigma_{S_{\mathrm{peak}}}$ and $\sigma_{\mathrm{S}}$ are independent, $R = \ln(S/S_{\mathrm{peak}})$ has a Gaussian distribution centred on zero whose rms is given by
\begin{equation}
\sigma_{\mathrm{R}} = \sqrt{ \left(\frac{\sigma_{\mathrm{S}}}{S} \right)^{2} + \left(\frac{\sigma_{S_{\mathrm{peak}}}}{S_{\mathrm{peak}}} \right)^{2}} \ .
\end{equation}
To detect an extended source at the 2$\sigma$ level, we require $R > 2 \sigma_{\mathrm{R}}$ or
\begin{equation}\label{eqn:indicator}
R > 2 \sqrt{ \left(\frac{\sigma_{\mathrm{S}}}{S} \right)^{2} + \left(\frac{\sigma_{S_{\mathrm{peak}}}}{S_{\mathrm{peak}}} \right)^{2}} \ .
\end{equation}
The probability of falsely classifying a point source as extended at the 2$\sigma$ level is 2.3 per cent. 

We used Equation~\ref{eqn:indicator} to separate point-like from extended sources.
$\sigma_{S_{\rm peak}}$ is taken as the sum in quadrature of the calibration error, the pixellation uncertainty \citep[see Appendices A and B of][]{Hales2012} and the local rms noise; the calibration error is set to $0.05~S_{\rm peak}$ (see Section~\ref{sec:Flux density calibration accuracy}) and the pixellation uncertainty to $0.01~S_{\rm peak}$. $\sigma_{S}$ is taken as the sum in quadrature of the calibration error and the local rms noise; the calibration error is set to $0.05~S$. At high SNR, where $\sigma_{\mathrm{S}}/S \approx 0.05$
and $\sigma_{\mathrm{S_{\mathrm{peak}}}}/S_{\mathrm{peak}} \approx 0.05$, $S/S_{\mathrm{peak}}$ must be greater than 1.15 for a source to be classified as extended.
The number of components classified as extended is 514 (17 per cent) in CDFS and 392 (19 per cent) in ELAIS-S1.

The deconvolved angular size for extended sources is given by \cite{Hales2014a}:
\begin{equation}\label{eqn:size}
\Theta = \sqrt{\theta_{\rm min}\theta_{\rm max} - B_{\rm maj}B_{\rm min}} \ .
\end{equation}
{\tt BLOBCAT} does not directly measure $\theta_{\rm min}$ or $\theta_{\rm max}$. By substituting for $\theta_{\rm min} \theta_{\rm max}$ using Equation~\ref{eqn:ratio},
\begin{equation}
\Theta = \sqrt{\left(\frac{S}{S_{\mathrm{peak}}}-1 \right) B_{\mathrm{maj}} B_{\mathrm{min}}} ~\mathrm{.}
\end{equation}
Following standard error propagation, the error on $\Theta$ is given by:
\begin{equation}
\sigma_\Theta = \frac{S}{S_{\rm peak}} \sqrt{\frac{B_{\rm maj}B_{\rm min}}{4(S/S_{\rm peak} - 1)} \left[\left(\frac{\sigma_{S_{\rm peak}}}{S_{\rm peak}}\right)^2 + \left(\frac{\sigma_S}{S}\right)^2\right]} \ .
\end{equation}

\subsection{Spectral Indices}\label{sec:sindex}

The wide bandwidth of the ATLAS observations provides enough
information to calculate the spectral indices over the mosaics. We
have measured the spectral indices of components where
the SNR was high enough to obtain a reliable measurement exclusively using the CABB data.

We created two separate mosaics of each field, one using the lower CABB subband data centred at $\nu_1=1.40\,$GHz 
and the other using the higher CABB subband data centred at $\nu_2=1.71\,$GHz, ensuring that both mosaics had 
the same resolution. We note that flagged frequency channels were accounted for in {\tt MIRIAD} when calculating 
the central frequencies of the two subbands.

To measure the spectral indices of isolated components, we ran {\tt BLOBCAT}
on the two CABB subband mosaics using a $5 \sigma$ detection limit. If a component was classified as extended in
the final catalogue, we used its integrated flux density at $\nu_{1}$ and $\nu_{2}$ as measured by {\tt BLOBCAT} to 
calculate its spectral index, otherwise we used its peak flux density. We did not measure the spectral indices of components 
detected below $5 \sigma$ at $\nu_{1}$ or $\nu_{2}$. 

We derived the spectral indices of complex components by
measuring their integrated flux densities in the two CABB subband images. The integrated flux density of a complex component
in each subband image was measured by summing the pixel values within its integration area (as measured in the 
raw {\tt BLOBCAT} catalogue described in Section~\ref{sec:srcext}), and dividing 
by the number of pixels per beam; the advantage of this method for measuring the spectral index is that it ensures that the flux density within the exact 
same region of the image is compared at each frequency, which is important for components 
spanning several beam areas. We did not attempt to measure the spectral indices of Gaussian components which were fitted to complex components.

The error on the spectral index was taken as
\begin{equation}
\label{eqn:err_alpha}
\sigma_{\alpha} = \sqrt{ \sigma_{\mathrm{\alpha, th}}^{2} + \sigma_{\mathrm{\alpha, cal}}^{2} } \ ,
\end{equation}
where $\sigma_{\mathrm{\alpha, th}}$ is the uncertainty on $\alpha$ due to the thermal noise and $\sigma_{\mathrm{\alpha, cal}}$
is the calibration error on $\alpha$. The noiselike uncertainty on $\alpha$ is given by
\begin{equation}
\label{eqn:err_alpha}
\sigma_{\mathrm{\alpha, th}} = \frac{\sqrt{\left[\frac{\sigma(\nu_1)}{S(\nu_1)}\right]^2+\left[\frac{\sigma(\nu_2)}{S(\nu_2)}\right]^2}}{\beta} \ ,
\end{equation}
where $S(\nu_1)$ and $S(\nu_2)$ are the flux densities at $\nu_1$ and $\nu_2$, $\sigma(\nu_1)$ and $\sigma(\nu_2)$ are the rms local noise at 
$\nu_1$ and $\nu_2$, and $\beta = {\rm ln}(\nu_2/\nu_1)$. For complex components spanning multiple beam areas, we multiplied
$\sigma(\nu_1)$ and $\sigma(\nu_2)$ by the square root of the integration area in units of the synthesised beam.

We have identified two main types of calibration errors contributing to $\sigma_{\mathrm{\alpha, cal}}$: uncorrelated errors in the primary 
beam model at $\nu_{1}$ and $\nu_{2}$, and uncorrelated errors in the antenna gain calibration at $\nu_{1}$ and $\nu_{2}$. Primary beam model
errors are expected to have the largest contribution close to the edges of the mosaics. As mentioned in Section~\ref{sec:srcext}, source extraction 
was limited to the region where the mosaicked primary beam response is above 40 per cent. The 40 per cent power level at $\nu_{2}$
corresponds to a distance from the pointing centre, $d$, of approximately 16~arcmin. In Section~\ref{sec:pbeam_effects}, we show that primary 
beam model errors cause $\alpha$ to flatten by no more than $\approx 0.1$ at $d = 20$~arcmin. We conclude that, for a component
located at any position within the mosaics, primary beam model errors introduce a spectral index error of at most $\approx 0.1$.

Given the very high degree of correlation ($\gtrapprox 95$ per cent) of gain calibration errors in the two CABB subbands, 
we have established that their contribution to $\sigma_{\mathrm{\alpha, cal}}$ is much less than 0.1. Since primary beam model errors are likely to 
be the dominant contribution to $\sigma_{\mathrm{\alpha, cal}}$, we set $\sigma_{\mathrm{\alpha, cal}} = 0.1$.

In total, we measured spectral indices for 1756 isolated components and 163 complex
components. Of the 1919 spectral index measurements, 344 have $\sigma_{\alpha}$ less than 0.2. We have included a column containing 
the spectral index in the final component catalogue for each field (see Table~\ref{tab:cat}). The spectral index properties 
of the ATLAS DR3 components are discussed further in Paper II (Banfield et al., in preparation).

\subsubsection{Verifying off-axis spectral index measurements}\label{sec:pbeam_effects}

In 2010, measurements of the primary beam for the new ATCA 16~cm receivers were carried out. The radial profile of the beam was measured along eight cuts at $45 \deg$ intervals, at 14 regularly-spaced frequencies across the band. The shape of the beam out to the first null was found to be close to circularly symmetric across the entire frequency range. A Gaussian fit to the beam was made at each frequency out to a level of $\sim 20 \%$ of the beam, using data from all angles. The product of the primary beam FWHM and the frequency was found to increase by 7 per cent between 1.3 and 1.8~GHz. This is probably due to a slight defocussing between these two frequencies. 

\begin{figure}
 \begin{center}
  \includegraphics[scale=0.30, angle=270]{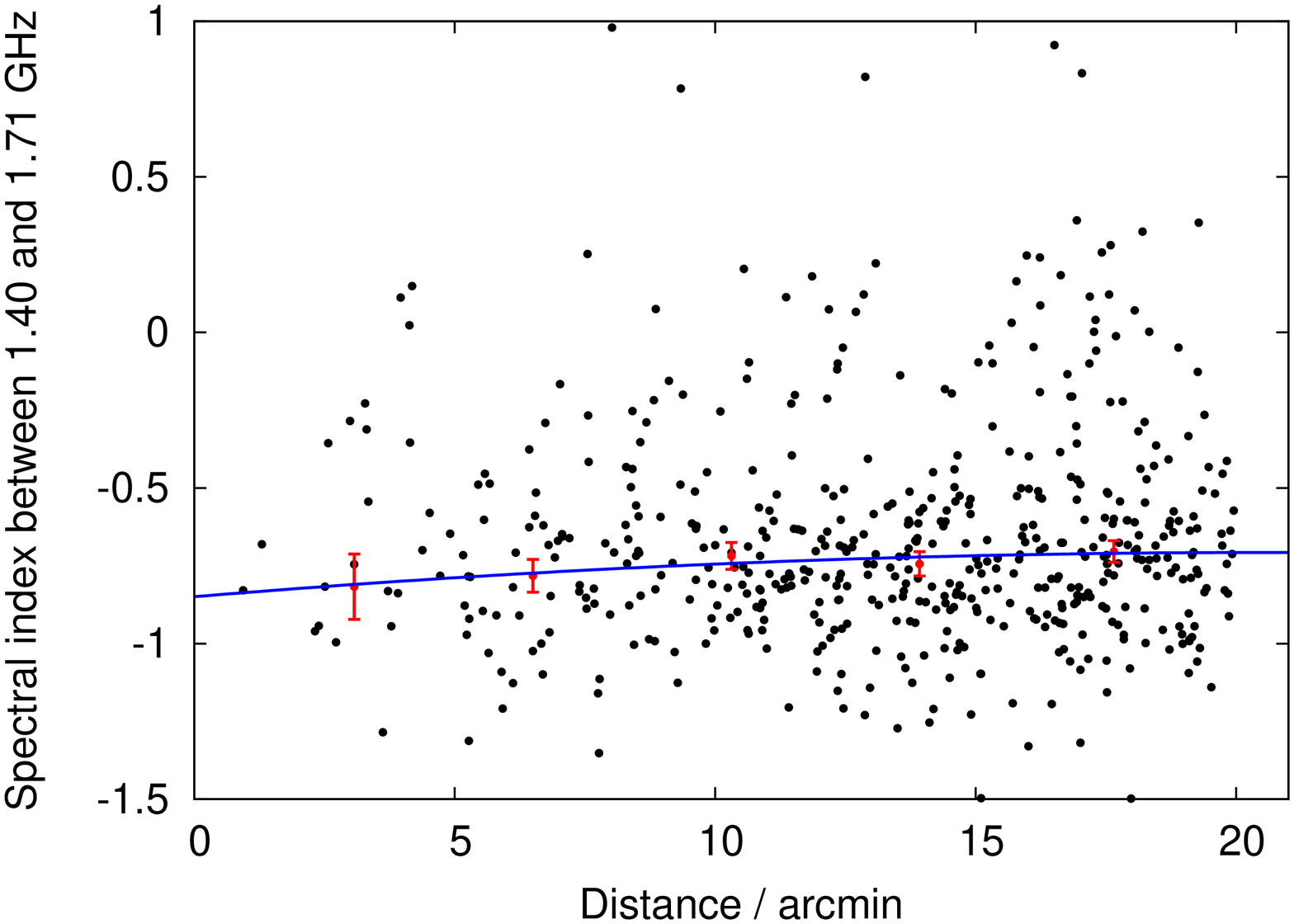}
  \includegraphics[scale=0.30, angle=270]{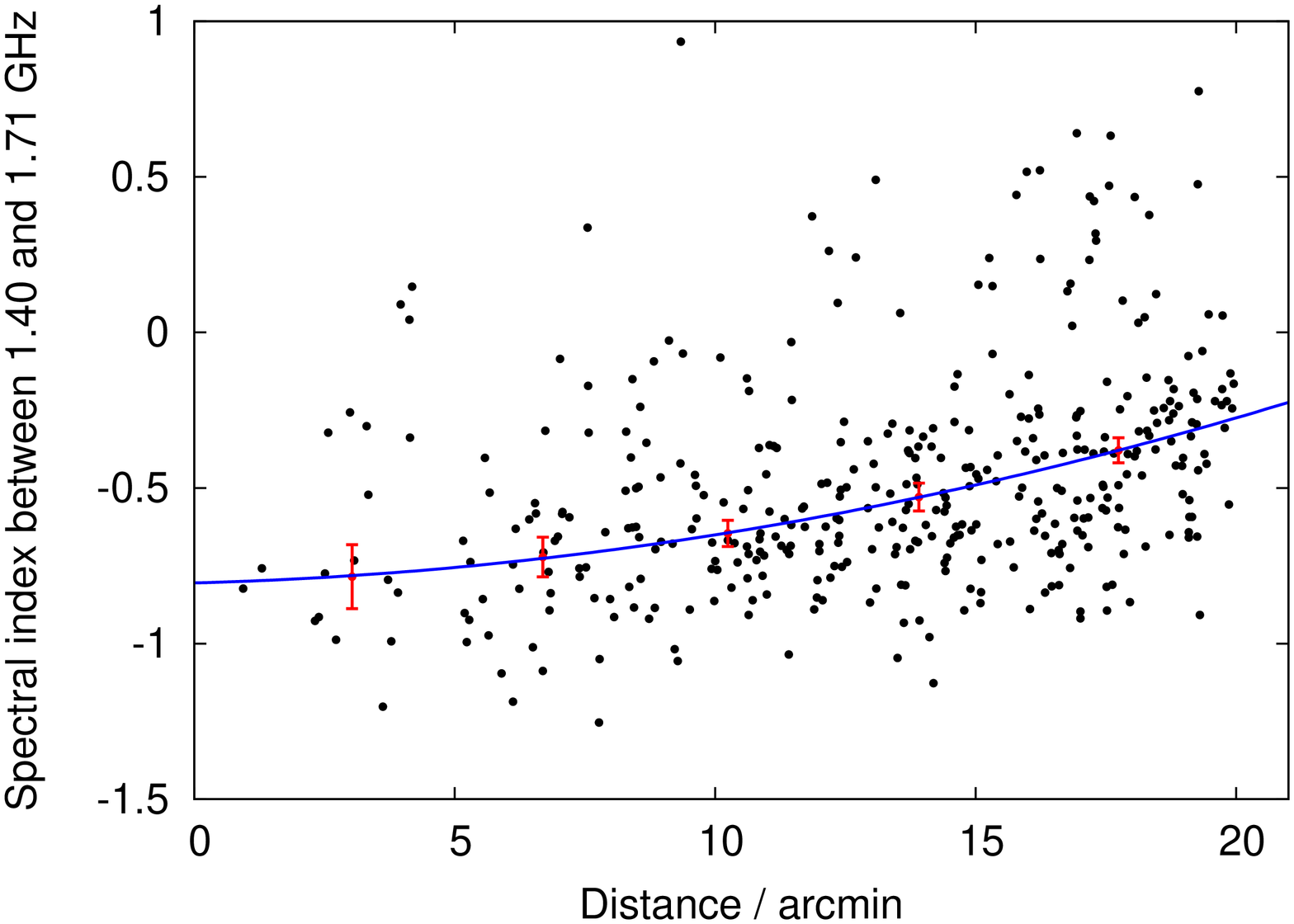}
  \caption{Spectral index between 1.40 and 1.71~GHz as a function of distance from the pointing centre for bright sources present in the CDFS and ELAIS-S1 pointings. The red points show median values in five distance bins, which are tabulated in Table~\ref{tab:var_vs_alpha}. The blue curve shows a quadratic fit to the median data points. Top: results using Gaussian primary beam fits for the new 16~cm CABB receivers between 1.1 and 3.1~GHz. Bottom: results obtained using an older primary beam model by \citet{wieringa1992}, where the beam FWHM varies as $\nu^{-1}$ between 1.3 and 1.8~GHz.}
  \label{fig:alpha_versus_dist_pbeam}
 \end{center}
\end{figure}

As mentioned in Section~\ref{mosaics}, we used these Gaussian primary beam fits for primary beam correction. A small error in the primary beam model may introduce a bias in the measured spectral indices away from the pointing centre. We therefore tested whether there was any systematic change in the spectral index with distance from the pointing centre. We applied the primary beam correction to each pointing in CDFS and ELAIS-S1 and restored them to a common resolution. We identified sources in the raw {\tt BLOBCAT} catalogue for the relevant field which were located within 20~arcmin from the pointing centre and with $\mathrm{SNR} > 200$. We measured the spectral indices of these sources by comparing the pixel values at their peak positions in the two CABB subband images. The top panel of Fig.~\ref{fig:alpha_versus_dist_pbeam} shows the measured spectral index as a function of the distance from the pointing centre, combining the results from all pointings. Given the SNR cut applied, the errors on the spectral indices are less than approximately 0.1. We measured the median spectral index in five distance bins (0--4, 4--8, 8--12, 12--16 and 16--20~arcmin) and fit a second order polymonial to the median data points. The median data points are shown in Table~\ref{tab:var_vs_alpha}.
There is no statistically significant change in the spectral index with distance from the pointing centre out to 20~arcmin, which corresponds to a power point of $\approx 0.25$ at 1.71~GHz. At $d = 20$~arcmin, we can be confident that errors in the primary beam model cause a flattening in the spectral index of no more than $\approx 0.1$.

To illustrate how sensitive the spectral index is to errors in the primary beam model, the bottom panel of Fig.~\ref{fig:alpha_versus_dist_pbeam} shows how the spectral index varies with distance from the pointing centre using an older primary beam model by \cite{wieringa1992}, where the beam FWHM varies as $\nu^{-1}$ between 1.3 and 1.8~GHz. In this case, there is a clear bias in the spectral index away from the pointing centre; the bias is $\approx 0.25$ at the half power point at 1.71~GHz.

\begin{table}
 \begin{center}
 \caption{Relationship between spectral index and distance from the pointing centre for bright sources in individual CDFS and ELAIS-S1 pointings. 
The first column gives the distance range and the second column the number of sources. The third column gives the median spectral index obtained using Gaussian primary beam fits for the new 16~cm CABB receivers between 1.1 and 3.1~GHz. The numbers in brackets are median spectral indices obtained using an older primary beam model by \citet{wieringa1992}, where the beam FWHM varies as $\nu^{-1}$ between 1.3 and 1.8~GHz.}
 \label{tab:var_vs_alpha}
 \begin{tabular}{@{} c c c }
 \hline
 $r$ & $N$ & $\alpha_{\mathrm{median}}$ \\
 \hline
0--4     &      17  &    $-0.82 \pm 0.11$ ($-0.79 \pm 0.10$) \\
4--8     &      59  &    $-0.78 \pm 0.05$ ($-0.72 \pm 0.06$) \\
8--12    &      105  &    $-0.72 \pm 0.04$ ($-0.65 \pm 0.04$) \\
12--16    &      159  &    $-0.74 \pm 0.04$ ($-0.53 \pm 0.05$) \\
16--20    &      184  &    $-0.70 \pm 0.04$ ($-0.38 \pm 0.04$) \\
 \hline
 \end{tabular}
 \end{center}
\end{table}

\subsection{Frequency Coverage}\label{sec:freq}

The combination of multiple epochs of ATLAS observations, the wider frequency coverage of the DR3 data and the amount of flagging resulted in the effective observing frequency changing slightly across the mosaics. We used the task {\tt linmos} to produce a mosaic giving the effective frequency across each field. In the final component catalogue, we have included the frequency at which the source's flux density is measured. The frequency typically lies in the range 1.45--1.50~GHz. 

We have also provided a column with the source's flux density at 1.4~GHz. This was derived using $\alpha^{1.71}_{1.40}$ if
$\sigma_{\alpha^{1.71}_{1.40}} \leq 0.2$, which is typically the case for sources with $S_{1.4} > 2$~mJy.
In their study of the sub-mJy radio population in the Lockman Hole, 
\cite{ibar2009} found no significant change in the median value of $\alpha^{1.40}_{0.61}$ as a function of $S_{1.4}$; 
$\alpha^{1.40}_{0.61}$ was found to be approximately --0.6 to --0.7 down to a flux density level of $S_{1.4} \gtrsim 100 \mu \mathrm{Jy}$.
For ATLAS sources with $\sigma_{\alpha^{1.71}_{1.40}} > 0.2$ or with no measured spectral indices, we therefore assumed that $\alpha^{1.71}_{1.40} = -0.7$
to obtain $S_{1.4}$.

\subsection{Calibration accuracy checks}\label{sec:Calibration accuracy checks}

\begin{figure}
 \begin{center}
  \includegraphics[scale=0.30, angle=270]{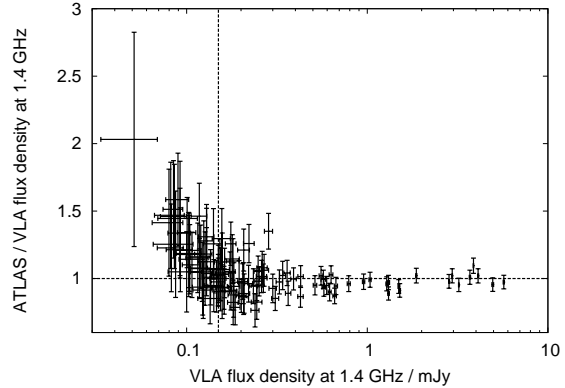}
  \caption{Ratio of the ATLAS to VLA flux density as a function of the VLA flux density for 112 unresolved sources in the eCDFS. The 
VLA flux densities were obtained from \citet{Miller2013}. The dashed horizontal line indicates equal flux density values
and the dashed vertical line indicates the flux density (0.15~mJy) above which the ATLAS flux densities are not 
considered to be affected by the Eddington bias.}
  \label{fig:atlas_vs_vla}
 \end{center}
\end{figure}

\begin{figure}
 \begin{center}
  \includegraphics[scale=0.50, trim=1cm 4cm 1cm 0cm, angle=270]{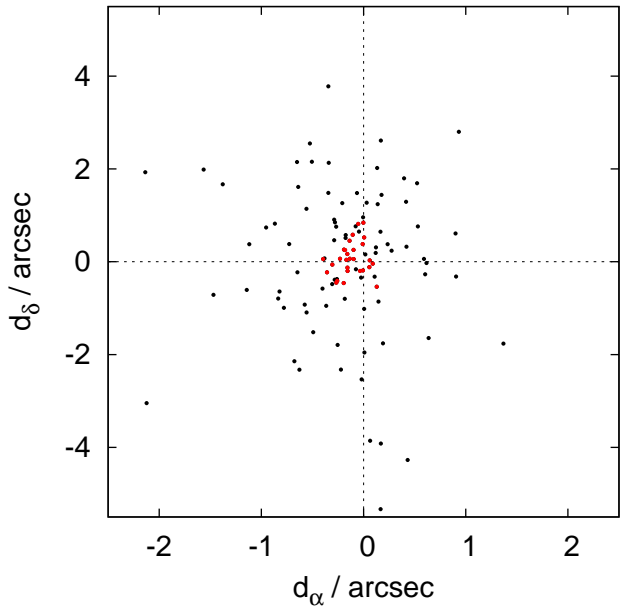}
  \caption{RA and Dec offsets between the ATLAS and VLA positions of 112 unresolved sources in the eCDFS.
Sources with ATCA SNRs $\geq 20$ are shown in red and the rest of the sources in black.}
  \label{fig:sep}
 \end{center}
\end{figure}

\cite{Miller2013} produced an image of the Extended Chandra Deep Field South (eCDFS) with the VLA at 1.4~GHz. Their
image covers an area of about a third of a square degree to a typical rms sensitivity of $7.4~\mu \mathrm{Jy} ~\mathrm{bm}^{-1}$
and has a resolution of 2.8 by 1.6~arcsec. In order to assess the accuracy of the ATCA flux densities and positions presented in this paper,
we matched our catalogue with the more sensitive and higher resolution catalogue by \citeauthor{Miller2013}. 

\subsubsection{Flux density calibration accuracy}\label{sec:Flux density calibration accuracy}

The flux density scale for ATCA is consistent with that of 
\cite{baars1977}, in use at the VLA, at the 1--2 per cent level over the range 1--10~GHz \citep{Reynolds1994}.
The VLA is more likely to resolve out part of the source flux density because of its smaller
beam size. In order to minimise discrepancies between the ATCA and VLA flux densities resulting from the smaller VLA beam
size, we only considered sources which were classified as point-like in both the ATCA and VLA catalogues.

Fig.~\ref{fig:atlas_vs_vla} shows how the ATCA and VLA integrated flux densities compare for all 112 unresolved sources in common
between the two surveys. 
There is generally excellent agreement between the flux densities. At the faint end, the ATCA flux densities tend to be 
systematically higher than the VLA flux densities. This is probably due, in part, to the Eddington bias \citep{eddington1913} causing 
the ATCA flux densities to be biased high close to the survey detection limit; the rms noise in the
eCDFS region of ATLAS is $\sim 3$ times higher than that in the VLA image. The flux density boosting resulting from Eddington bias 
only depends on the SNR and the source count slope \citep{hogg1998}. \cite{Hales2014a} used the 1.4~GHz source count fit by \cite{hopkins2003}
to evaluate the flux density boosting for a $5 \sigma$ source as a function of $S_{1.4}$; the bias was estimated to be $\approx 10$ per cent 
at $S_{1.4} = 0.1~\mathrm{mJy}$. This is not sufficient to explain the observed discrepancy between the ATCA and VLA flux densities at the faint end.
Another potential cause of the discrepancy is missing extended flux density in the VLA image. 

For sources with $S_{\rm VLA} > 0.15$~mJy, the ATCA flux densities 
do not appear to be affected by the Eddington bias. For these 74 sources, the ATCA flux densities are, on average, consistent 
with the VLA flux densities at the $\approx 2$ per cent level: the mean value of $S_{\rm ATCA}/S_{\rm VLA}$ is $0.986 \pm 0.013$ 
and the median value is $0.974 \pm 0.016$.

At high SNRs, calibration errors will become the main source of uncertainty.
Of the 112 ATCA-VLA sources, 29 have ATCA SNRs $\geq 20$. For these 29 sources, the standard deviation of $S_{\rm ATCA}/S_{\rm VLA}$
is 0.047. We therefore set the ATCA flux density calibration error to 5 per cent. This is a conservative estimate 
of the flux density calibration error since it does not account for the scatter in $S_{\rm ATCA}/S_{\rm VLA}$ introduced by
errors on the VLA flux densities.

\subsubsection{Positional accuracy}\label{Positional accuracy}

We have estimated the calibration errors $\sigma_{\mathrm{\alpha,cal}}$ and $\sigma_{\mathrm{\delta,cal}}$ in RA and Dec 
by comparing the ATCA positions with the VLA positions. Fig.~\ref{fig:sep} shows the RA offset, $d_{\alpha}$, and the Dec offset, 
$d_{\delta}$, for the 112 unresolved sources in common between the two surveys. Sources with ATCA SNRs $\geq 20$, for which calibration 
errors dominate the position uncertainties, are shown in red.

The positional uncertainties in RA and Dec resulting from phase errors can be expressed as $b ~\Theta_{\alpha}$
and $b ~\Theta_{\delta}$, where $b$ is a constant, and $\Theta_{\alpha/\delta}$ is the projected resolution 
in the RA/Dec direction. For the 29 sources with ATCA SNRs $\geq 20$, the standard deviation of $d_{\alpha}$ is 
0.13~arcsec and the standard deviation of $d_{\delta}$ is 0.36~arcsec, indicating that $b \approx 0.021$.

To test for systematic errors, we calculated the mean 
offsets in RA and Dec. The mean value of $d_{\alpha}$ is ($-0.12 \pm 0.02$)~arcsec and the mean value
of $d_{\delta}$ is ($0.07 \pm 0.07$)~arcsec. To account for the systematic offset in RA, we increased $\sigma_{\mathrm{\alpha,cal}}$ by
0.12~arcsec. In CDFS, we set $\sigma_{\mathrm{\alpha,cal}} = b ~\Theta_{\alpha} + 0.12~\mathrm{arcsec} = 0.26~\mathrm{arcsec}$.
Since the mean value of $d_{\delta}$ is consistent with zero at the 1$\sigma$ level, we assumed no systematic offset 
in Dec and set $\sigma_{\mathrm{\delta,cal}} = b ~\Theta_{\delta} = 0.33~\mathrm{arcsec}$. 

It was not possible
to carry out a similar analysis in ELAIS-S1 because no high resolution radio data were available to perform a 
comparison. Assuming no systematic errors in RA and Dec, and using the same value of $b$ as for CDFS, 
$\sigma_{\mathrm{\alpha,cal}} = b ~\Theta_{\alpha} = 0.16$~arcsec and $\sigma_{\mathrm{\delta,cal}} = b ~\Theta_{\delta} = 0.25$~arcsec.

We set the total position errors $\sigma_{\mathrm{\alpha}}$ and $\sigma_{\mathrm{\delta}}$ in RA and Dec to
\begin{subequations}
\begin{equation}
\sigma_{\mathrm{\alpha}}= \sqrt{ \sigma_{\mathrm{\alpha,cal}}^{2} + \sigma_{\mathrm{\alpha,th}}^{2} }
\end{equation}    
\begin{equation}
\sigma_{\mathrm{\delta}}= \sqrt{ \sigma_{\mathrm{\delta,cal}}^{2} + \sigma_{\mathrm{\delta,th}}^{2} } ~,
\end{equation}
\end{subequations}
where $\sigma_{\mathrm{\alpha,th}}$ and $\sigma_{\mathrm{\delta,th}}$ are the position errors in RA and Dec due to
the significance of the component detection. \textsc{blobcat} (Hales et al. 2012) estimates $\sigma_{\mathrm{\alpha,th}}$ and 
$\sigma_{\mathrm{\delta,th}}$ as
\begin{subequations}
\begin{equation}
\sigma_{\mathrm{\alpha,th}}= \frac{1}{1.4 A} \Theta_{\alpha}
\end{equation}    
\begin{equation}
\sigma_{\mathrm{\delta,th}}= \frac{1}{1.4 A} \Theta_{\delta} ~,
\end{equation}
\end{subequations}
where $A$ is the SNR.

\subsection{Component Catalogue}\label{sec:cat}
We have combined the source component catalogues for both the CDFS and ELAIS-S1 into one component source catalogue for ATLAS.  Table \ref{tab:cat} lists the first 14 components in the catalogue while the full catalogue can be obtained from the supplementary material.  The columns of the component catalogue are defined as follows:

{\it Column (1)} -- component number. A prefix indicates the field in which the component lies: `CI' and `EI' stand for Stokes $I$ component in CDFS and ELAIS-S1, respectively. Where multiple Gaussians were fitted to complex components, the component number is appended by the letter `C' followed by the Gaussian number.

{\it Column (2)} -- component IAU name given by ATLAS3 Jhhmmss.s-ddmmssC, where the letter `C' stands for component in Stokes $I$. Norris et al, in preparation compiles the ATLAS DR3 source catalogue by combining the source components.

{\it Columns (3) and (4)} -- intensity weighted centroid position: RA (J2000), $\alpha$, in hours:minutes:seconds, and Dec. (J2000), $\delta$, in degrees:minutes:seconds.

{\it Columns (5) and (6)} -- intensity weighted centroid position: RA (J2000), $\alpha$, and Dec. (J2000), $\delta$, in deg.

{\it Columns (7) and (8)} -- error on centroid position in RA, $\sigma_{\alpha}$, and in Dec., $\sigma_{\delta}$, in arcsec. These were derived as described in Section~\ref{Positional accuracy}.

{\it Column (9)} -- local rms noise level, $\sigma_{\rm local}$, in mJy beam$^{-1}$.

{\it Column (10)} -- bandwidth smearing correction, $b$ (see Section~\ref{sec:bws}).

{\it Column (11)} -- effective frequency, $\nu_{\rm obs}$, in~GHz, at which the source was measured in the mosaic (see Section~\ref{sec:freq}).

{\it Columns (12) and (13)} -- peak flux density, $S_{\rm peak}$, and associated error, $\sigma_{S_{\rm peak}}$, at $\nu_{\rm obs}$, in mJy beam$^{-1}$. $S_{\rm peak}$ has been corrected for peak bias and bandwidth smearing in {\tt BLOBCAT}. $\sigma_{S_{\rm peak}}$ is derived as described in Section~\ref{sec:res}. $\nu_{\rm obs}$ is provided in column (20).

{\it Columns (14) and (15)} -- integrated flux density, $S$, and associated error, $\sigma_{S}$, at $\nu_{\rm obs}$, in mJy. $\sigma_{S}$ is derived as described in Section~\ref{sec:res}. $S$ is set to $S_{\rm peak}$ and $\sigma_{S}$ to $\sigma_{S_{\rm peak}}$ for sources classified as point-like. 

{\it Columns (16) and (17)} -- peak flux density, $S_{\rm peak, 1.4}$, and associated error, $\sigma_{S_{\rm peak, 1.4}}$, at 1.4~GHz, in mJy beam$^{-1}$. $S_{\rm peak, 1.4}$ was derived from $S_{\rm peak}$
as described in Section~\ref{sec:freq}.

{\it Columns (18) and (19)} -- integrated flux density, $S_{1.4}$, and associated error, $\sigma_{S_{1.4}}$, at 1.4~GHz, in mJy. $S_{1.4}$ was derived from $S$ as described in Section~\ref{sec:freq}.

{\it Columns (20) and (21)} -- deconvolved angular size, $\Theta$, and associated error, $\sigma_{\Theta}$, in arcsec, for sources classified as extended. $\Theta$ and $\sigma_{\Theta}$ are set to $-999$ for point sources. 

{\it Column (22)} -- component type, $t$: point-like (P) or extended (E) (see Section~\ref{sec:res}).

{\it Columns (23) and (24)} -- spectral index between 1.40 and 1.71~GHz, $\alpha^{1.71}_{1.40}$, and associated error, $\sigma_{\alpha^{1.71}_{1.40}}$, as measured in Section~\ref{sec:sindex}. The spectral index is set to $-999$ if the SNR was too low for it to be measured.

{\it Column (25)} -- field identifier.

\begin{landscape}
\begin{table}
\caption{Catalogue entries for the first 14 components in the ATLAS $1.4\,$GHz component catalogue.  The columns are defined in Section~\ref{sec:cat}.  The full catalogue can be found in the supplementary material using the same column format.}\label{tab:cat}
 \begin{tabular}{rrrrrrrrrrrr}
 \hline
Component number & IAU Name & $\alpha$ & $\delta$ & $\alpha$ & $\delta$ & $\sigma_{\alpha}$ & $\sigma_{\delta}$ & $\sigma_{\rm local}$ & $b$ & $\nu_{\rm obs}$ & $S_{\rm peak}$ \\
  &  & & & (deg) & (deg) & (arcsec) & (arcsec) & (mJy bm$^{-1}$) & & (GHz) & (mJy bm$^{-1}$) \\
 (1) & (2) & (3) & (4) & (5) & (6) & (7) & (8) & (9) & (10) & (11) & (12) \\
 \hline
EI0001 & ATLAS3 J002925.7-440256C & 00:29:25.73 & -44:02:56.6 & 7.357191 & -44.049053 & 0.44 & 0.69 & 0.052 & 1.08 & 1.439 & 0.751 \\
EI0002 & ATLAS3 J002926.7-440016C & 00:29:26.79 & -44:00:16.5 & 7.361610 & -44.004578 & 1.07 & 1.65 & 0.051 & 1.08 & 1.440 & 0.288 \\
EI0003 & ATLAS3 J002933.7-440118C & 00:29:33.70 & -44:01:18.4 & 7.390433 & -44.021768 & 0.79 & 1.22 & 0.046 & 1.08 & 1.447 & 0.349 \\
EI0004 & ATLAS3 J002938.1-432946C & 00:29:38.17 & -43:29:46.2 & 7.409025 & -43.496162 & 0.34 & 0.52 & 0.045 & 1.09 & 1.446 & 0.904 \\
EI0005 & ATLAS3 J002938.9-440031C & 00:29:38.91 & -44:00:31.2 & 7.412115 & -44.008663 & 0.70 & 1.08 & 0.043 & 1.07 & 1.452 & 0.367 \\
EI0006 & ATLAS3 J002940.1-440308C & 00:29:40.19 & -44:03:08.7 & 7.417454 & -44.052408 & 0.60 & 0.92 & 0.042 & 1.07 & 1.452 & 0.434 \\
EI0007 & ATLAS3 J002941.3-435334C & 00:29:41.34 & -43:53:34.0 & 7.422268 & -43.892779 & 0.61 & 0.95 & 0.043 & 1.08 & 1.440 & 0.430 \\
EI0008 & ATLAS3 J002941.8-440714C & 00:29:41.80 & -44:07:14.1 & 7.424153 & -44.120575 & 0.99 & 1.53 & 0.042 & 1.08 & 1.447 & 0.252 \\
EI0009 & ATLAS3 J002943.1-440813C & 00:29:43.11 & -44:08:13.3 & 7.429629 & -44.137028 & 0.77 & 1.19 & 0.042 & 1.08 & 1.446 & 0.333 \\
EI0010 & ATLAS3 J002944.3-433629C & 00:29:44.30 & -43:36:29.0 & 7.434567 & -43.608068 & 0.63 & 0.98 & 0.049 & 1.09 & 1.435 & 0.480 \\
EI0011C1 & ATLAS3 J002945.3-432148C & 00:29:45.34 & -43:21:48.0 & 7.438908 & -43.363342 & 0.16 & 0.26 & 0.042 & 1.08 & 1.444 & 6.506 \\
EI0011C2 & ATLAS3 J002946.1-432148C & 00:29:46.17 & -43:21:48.7 & 7.442358 & -43.363537 & 0.16 & 0.26 & 0.042 & 1.08 & 1.445 & 7.060 \\
EI0012 & ATLAS3 J002946.3-440724C & 00:29:46.33 & -44:07:24.0 & 7.443055 & -44.123333 & 0.82 & 1.27 & 0.040 & 1.07 & 1.450 & 0.289 \\
EI0013 & ATLAS3 J002948.6-435618C & 00:29:48.66 & -43:56:18.6 & 7.452745 & -43.938512 & 0.71 & 1.10 & 0.037 & 1.07 & 1.454 & 0.315 \\
\hline
\end{tabular}
\end{table}
\begin{table}
\contcaption{}
 \begin{tabular}{rrrrrrrrrrrrr}
 \hline
$\sigma_{S_{\rm peak}}$ & $S$ & $\sigma_{\rm S}$  & $S_{\rm peak, 1.4}$ & $\sigma_{S_{\rm peak, 1.4}}$ & $S_{1.4}$ & $\sigma_{S_{1.4}}$ & $\Theta$ & $\sigma_{\Theta}$ & $t$ & $\alpha^{1.71}_{1.40}$ & $\sigma_{\alpha^{1.71}_{1.40}}$ & Field \\
(mJy bm$^{-1}$) & (mJy) & (mJy) & (mJy bm$^{-1}$) & (mJy bm$^{-1}$) & (mJy) & (mJy) & (arcsec) & (arcsec) & & & & \\
 (13) & (14) & (15) & (16) & (17) & (18) & (19) & (20) & (21) & (22) & (23) & (24) & (25) \\
 \hline
0.068 & 0.751 & 0.068 & 0.766 & 0.070 & 0.766 & 0.070 & -999.0 & -999.0 & P & -0.83 & 1.01 & ELAIS-S1 \\
0.057 & 0.288 & 0.057 & 0.294 & 0.059 & 0.294 & 0.059 & -999.0 & -999.0 & P & -999.00 & -999.00 & ELAIS-S1 \\
0.052 & 0.349 & 0.052 & 0.357 & 0.054 & 0.357 & 0.054 & -999.0 & -999.0 & P & -999.00 & -999.00 & ELAIS-S1 \\
0.067 & 0.904 & 0.067 & 0.924 & 0.069 & 0.924 & 0.069 & -999.0 & -999.0 & P & -2.72 & 0.93 & ELAIS-S1 \\
0.049 & 0.367 & 0.049 & 0.377 & 0.051 & 0.377 & 0.051 & -999.0 & -999.0 & P & -999.00 & -999.00 & ELAIS-S1 \\
0.051 & 0.434 & 0.051 & 0.445 & 0.052 & 0.445 & 0.052 & -999.0 & -999.0 & P & -999.00 & -999.00 & ELAIS-S1 \\
0.051 & 0.430 & 0.051 & 0.438 & 0.052 & 0.438 & 0.052 & -999.0 & -999.0 & P & -999.00 & -999.00 & ELAIS-S1 \\
0.047 & 0.252 & 0.047 & 0.258 & 0.048 & 0.258 & 0.048 & -999.0 & -999.0 & P & -999.00 & -999.00 & ELAIS-S1 \\
0.049 & 0.333 & 0.049 & 0.340 & 0.050 & 0.340 & 0.050 & -999.0 & -999.0 & P & -999.00 & -999.00 & ELAIS-S1 \\
0.059 & 0.480 & 0.059 & 0.489 & 0.060 & 0.489 & 0.060 & -999.0 & -999.0 & P & -999.00 & -999.00 & ELAIS-S1 \\
0.330 & 7.983 & 0.406 & 6.703 & 0.341 & 8.224 & 0.420 & 4.6 & 0.9 & E & -0.96 & 0.15 & ELAIS-S1 \\
0.357 & 8.632 & 0.438 & 7.278 & 0.370 & 8.899 & 0.454 & 4.5 & 0.9 & E & -0.96 & 0.15 & ELAIS-S1 \\
0.045 & 0.289 & 0.045 & 0.296 & 0.046 & 0.296 & 0.046 & -999.0 & -999.0 & P & -999.00 & -999.00 & ELAIS-S1 \\
0.043 & 0.315 & 0.043 & 0.323 & 0.044 & 0.323 & 0.044 & -999.0 & -999.0 & P & -999.00 & -999.00 & ELAIS-S1 \\
\hline
\end{tabular}
\end{table}
\end{landscape}

\section{Conclusion}\label{Summary}

We present images and a component catalogue from a deep, wideband, radio continuum survey with ATCA. The third ATLAS data release combines observations taken between 2002 and 2010 of an area coincident with the CDFS and ELAIS-S1, two of the best-studied regions of the sky at all wavelengths. The survey covers a total area of 6.3~$\mathrm{deg}^{2}$ to a typical rms noise level of $15\,\mu$Jy beam$^{-1}$ at 1.4~GHz. Various array configurations were employed to maximise the \textit{uv} coverage, resulting in a resolution of 16 by 7~arcsec in CDFS and 12 by 8~arcsec in ELAIS-S1. ATLAS is among the deepest and widest radio surveys to date and is being used as a pilot survey for EMU, which will cover the whole Southern Sky to approximately the same depth as ATLAS, at a similar resolution and frequency.

In Paper II, we present the first results from the survey, including the deep 1.4-GHz source counts. Here, we have concentrated on developing techniques for producing and analysing the radio maps to enable maximum scientific return from the survey. In particular, we have:

\begin{enumerate}

\item[(1)] developed strategies to automatically flag data taken with the new CABB correlator between 1.3 and 1.8~GHz, and calibrate the data while accounting for frequency-dependent variations in the gains;

\item[(2)] used a variety of techniques, such as multi-frequency CLEAN, self-calibration and wideband primary beam correction, to produce images with high dynamic range and fidelity;

\item[(3)] run the {\tt BLOBCAT} source finder on our maps while applying corrections for bandwidth smearing and peak bias to the source flux densities, resulting in a total of 5118 components above $5 \sigma$ in the two fields.

\item[(4)] used the wide bandwidth of our observations to measure the spectral indices across the CABB band (1.3--1.8~GHz) of $\sim 2000$ of the brightest components in the fields. Analysis of the spectral index results will be presented in Paper II.

\end{enumerate}

\section*{Acknowledgments}

We thank the referee, Jim Condon, for helpful comments. We would also like to thank the following people for discussions on calibration and imaging of ATCA CABB data: K.~Bannister, S.~Brown, T.~Cornwell, I.~Feain, N.~McClure-Griffiths, D.~Schnitzeler, S.~O'Sullivan and M.~Wieringa. TMOF acknowledges support from an ARC Super Science Fellowship. JKB acknowledges funding from the Australian Research Council Centre of Excellence for All-sky Astrophysics (CAASTRO), through project number CE110001020. NS acknowledges support from an ARC Future Fellowship. The Australia Telescope Compact Array is part of the Australia Telescope, which is funded by the Commonwealth of Australia for operation as a National Facility managed by CSIRO. The National Radio Astronomy Observatory is a facility of the National Science Foundation operated under cooperative agreement by Associated Universities, Inc.

\newcommand{\aj}{AJ} 
\newcommand{\apj}{ApJ}
\newcommand{\mnras}{MNRAS} 
\newcommand{\pasa}{PASA}
\newcommand{\aap}{A\&A}
\newcommand{\araa}{ARA\&A}
\newcommand{\aaps}{A\&AS}
\newcommand{\aujpa}{AuJPA}
\newcommand{\apjs}{ApJS}

\bibliographystyle{mn2e}
\bibliography{apjmnemonic,bibtex-references}


\appendix\section{Observational information on ATLAS fields}\label{Observational information on ATLAS fields}


\begin{table*}
 \centering
  \caption{Observational information on the two ATLAS fields including observing dates, ATCA array configurations, and net integration times after calibration and flagging.}\label{Tab:obsinfo}
  \begin{tabular}{lllcc}
  \hline
 Field & Project & Date & Configuration & Integration \\
 & ID & & & Time (h)\\
 \hline
 CDFS & C1035$^{a,c,d}$ & 2002 Apr 4--7, 10, 12--13 & 6A & 72.9\\
 & & 2002 Aug 23--24, 27--29 & 6C & 29.6\\
 & C1241$^{a,c,d}$ & 2004 Jan 7--8, 12 & 6A & 23.9\\
 & & 2004 Feb 3--5 & 6B & 24.7\\
 & & 2004 Jun 6, 8--12 & 750D & 37.4\\
 & & 2004 Nov 24--30 & 6D & 50.4\\
 & & 2004 Dec 28--30 & 1.5D & 22.6\\
 & C1241$^{c,d}$  & 2005 Jan 7--8, 18--19, 23 & 750B & 31.9\\
 & & 2005 Apr 9--10 & 6A & 18.5\\
 & & 2005 Apr 14  & 1.5A & 8.9\\
 & & 2005 Apr 22; 2005 May 2 & 750A & 15.0\\
 & & 2005 Jun 1, 10 & EW367 & 11.7\\
 & & 2005 Jun 25--26 & 6B & 18.1\\
 & & 2005 Dec 6 & 6A & 8.7\\
 & & 2006 Mar 23--24, 27 & 6C & 23.0\\
 & C1967$^d$ & 2009 Jun 19--21, Dec 19--31, 2010 Jan 1--3, Apr 1--6, 12, 14--18 & 6A & 148.1\\
  & & 2009 Aug 13, 19, 21--27 & 6D & 76.6 \\
  & & 2009 Nov 10--18, 20--29 & 6B & 161.0 \\
  & & 2010 May 1--3, 20--24, Jun 1, 5--9, 11--14, 27 & 6C & 40.5 \\
  \hline
ELAIS-S1 & C1241$^{b,c,d}$ & 2004 Jan 9--11 & 6A & 24.6\\
 & & 2004 Jan 30, Feb 1 & 6B & 18.6\\
 & & 2004 Dec 19, 27, 2005 Jan 1--3 & 1.5D & 40.2\\
 & & 2005 Jan 9--11, 20--22 & 750B & 50.0\\
 & & 2005 Mar 25, Apr 8, 11 & 6A & 27.2\\
 & & 2005 Apr 24, 26, 30, May 1 & 750A & 34.3\\
 & & 2005 June 8, 9 & EW367 & 18.3\\
 & & 2005 June 19, 24 & 6B & 18.5\\
 & C1967$^d$ & 2009 Dec 19--31, 2010 Jan 1--3, Apr 1--6, 12, 14--18 & 6A & 68.6\\
 & & 2009 Aug 13, 19, 21--27 & 6D & 87.9 \\
 & & 2009 Nov 10--18, 20--29 & 6B & 9.8 \\
 & & 2010 Apr 19 & 6C & 10.6 \\
\hline
& & $^a$ Data presented in \citet{Norris2006}.\\
& & $^b$ Data presented in \citet{Middelberg2008}.\\
& & $^c$ Data presented in \citet{Hales2014a}.\\
& & $^d$ Data presented in this work.
\end{tabular}
\end{table*}

\begin{table*}
 \centering
  \caption{Coordinates of mosaic pointings in the two ATLAS fields.}
\label{Tab:pntinfo}
 \begin{tabular}{lll}
 \hline
 Pointing & R.A. (J2000) & Decl. (J2000)\\
 \hline
CDFS\\
1 & 03:28:47.33 & $-$28:38:37.98 \\
1a & 03:27:18.36 & $-$28:38:31.14 \\
2 & 03:28:03.89 & $-$28:21:46.74 \\
3 & 03:28:48.48 & $-$28:05:05.58 \\
3a & 03:27:18.36 & $-$28:05:05.58 \\
4 & 03:28:05.26 & $-$27:48:14.34 \\
5 & 03:28:49.61 & $-$27:31:32.82 \\
5a & 03:27:18.36 & $-$27:31:32.82 \\
10 & 03:30:16.97 & $-$27:31:40.02 \\
11 & 03:29:32.83 & $-$27:48:22.98 \\
12 & 03:30:16.30 & $-$28:05:12.42 \\
13 & 03:29:31.92 & $-$28:21:55.74 \\
14 & 03:30:15.60 & $-$28:38:44.82 \\
15 & 03:31:43.87 & $-$28:38:48.42 \\
16 & 03:30:59.95 & $-$28:22:00.78 \\
27 & 03:32:27.99 & $-$28:22:02.58 \\
28 & 03:33:12.12 & $-$28:38:48.42 \\
29 & 03:34:40.39 & $-$28:38:44.82 \\
30 & 03:33:56.02 & $-$28:22:00.78 \\
31 & 03:34:39.70 & $-$28:05:12.42 \\
33 & 03:34:39.03 & $-$27:31:40.02 \\
41 & 03:32:28.00 & $-$27:48:30.00 \\
42 & 03:31:20.17 & $-$27:48:30.00 \\
43 & 03:31:54.08 & $-$28:01:29.44 \\
44 & 03:33:01.92 & $-$28:01:29.44 \\
45 & 03:33:35.83 & $-$27:48:30.00 \\
46 & 03:33:01.92 & $-$27:35:30.56 \\
47 & 03:31:54.08 & $-$27:35:30.56 \\
\hline
ELAIS-S1\\
1 & 00:32:03.55 & $-$43:44:51.24\\
2 & 00:31:10.95 & $-$43:27:59.64\\
3 & 00:32:05.04 & $-$43:11:18.84\\
4 & 00:33:51.29 & $-$43:11:24.96\\
5 & 00:32:57.67 & $-$43:28:09.00 \\
6 & 00:33:50.79 & $-$43:44:57.36 \\
7 & 00:35:38.02 & $-$43:44:57.36\\
8 & 00:34:44.40 & $-$43:28:11.88 \\
9 & 00:35:37.51 & $-$43:11:24.96 \\
10 & 00:37:23.76 & $-$43:11:18.84\\
11 & 00:36:31.13 & $-$43:28:09.00 \\
12 & 00:37:25.25 & $-$43:44:51.24 \\
13 & 00:36:31.13 & $-$44:01:42.84 \\
14 & 00:37:25.25 & $-$44:18:34.44\\
15 & 00:35:38.02 & $-$44:18:34.44\\
16 & 00:34:44.40 & $-$44:01:42.84\\
17 & 00:32:57.67 & $-$44:01:42.84\\
18 & 00:33:50.79 & $-$44:18:34.44 \\
19 & 00:32:03.55 & $-$44:18:34.44\\
20 & 00:31:10.95 & $-$44:01:42.84\\
     \hline
\end{tabular}
\end{table*}
\label{lastpage}

\end{document}